\documentclass[useAMS,usenatbib]{mn2e}
\usepackage{graphicx}
\usepackage{amsmath,amsfonts,amssymb}
\usepackage{wrapfig}
\usepackage{epsfig}
\usepackage{psfrag}
\usepackage{pstricks,pst-plot} 

\begin{document}
%%%%%%%%%%%%%%%%%%%%%%%%%%%%%%%%%%%%%%%%%%%%%%%%

\title[Recycled radio pulsars in globular clusters]
{Luminosities of recycled radio pulsars in globular clusters}

\author[Bagchi, Lorimer \& Chennamangalam]
{\parbox[t]{\textwidth}{Manjari Bagchi$^{1}$\thanks{Email: Manjari.Bagchi@mail.wvu.edu}, D.R.~Lorimer$^{1,2}$ and Jayanth Chennamangalam$^{1}$}\\
\vspace*{3pt} \\
\\ $^1$ Department of Physics, 210 Hodges Hall, West Virginia University, Morgantown, WV 26506, USA
\\ $^2$ NRAO, Green Bank Observatory, PO Box 2, Green Bank, WV 24944, USA
}
\maketitle

\begin{abstract}
Using Monte Carlo simulations, we model the luminosity distribution of
recycled pulsars in globular clusters as the brighter, observable part
of an intrinsic distribution and find that the observed luminosities
can be reproduced using either log-normal or power-law distributions
as the underlying luminosity function. For both distributions, a wide
range of model parameters provide an acceptable match to the observed
sample, with the log-normal function providing statistically better
agreement in general than the power-law models. Moreover, the power-law
models predict a parent population size that is a factor of between
two and ten times higher than for the log-normal models. We note that
the log-normal luminosity distribution found for the normal pulsar
population by Faucher-Gigu\`ere and Kaspi is consistent with the
observed luminosities of globular cluster pulsars. For Terzan~5, our
simulations show that the sample of detectable radio pulsars, and the
diffuse radio flux measurement, can be explained using the log-normal
luminosity law with a parent population of $\sim 150$ pulsars. Measurements 
of diffuse gamma-ray fluxes for several clusters can be explained by both power-law and log-normal models, with the
log-normal distributions again providing a better match in general.
In contrast to previous studies, we do not find any strong evidence
for a correlation between the number of pulsars inferred in globular
clusters and globular cluster parameters including metallicity and
stellar encounter rate.
\end{abstract}

\begin{keywords}
{stars: neutron --- pulsars: general --- methods: numerical --- methods: statistical --- globular clusters: general --- globular clusters: individual (Terzan 5)}
\end{keywords}

\section{Introduction} 
\label{lb:introductions} 

The first millisecond radio pulsar in a globular cluster (GC) was
discovered by \citet{lbmkb87}, shortly after earlier predictions that
the putative progenitors of millisecond pulsars (MSP) are low-mass
X-ray binaries \citep{acrs82} which are known to be present in GCs
\citep{katz75}. Inspired by this discovery, a large number of
sensitive pulsar searches have been performed over the years resulting
in the currently observed population of 143 radio pulsars in 27
GCs\footnote{For a complete list, see
  http://www.naic.edu/$\sim$pfreire/GCpsr.html.}. GCs have some
physical properties which are different than those of the galactic
disk. Examples are extremely high stellar density and high abundance
of metal poor population II stars indicating that they were born in
the early phase of the Galaxy's formation. These facts lead naturally
to the question whether the population of radio pulsars in GCs is
different to their counterparts in the Galactic disk. A number of
differences are already well-known. In particular, the high abundance
of MSPs both in eccentric binary systems and as isolated objects.
These phenomena can be explained as the results of two-body or
three-body stellar interactions in the dense stellar environment of
GCs \citep{rkb87, vhpr87, ihrbf08, br09}. On the other hand, the
luminosity of a pulsar is a more fundamental property as it can in
principle be linked to the pulsar emission mechanism. It is therefore
important to establish whether there is any evidence for a different
luminosity function of MSPs in the disk compared to those in GCs.

There are two main ways to determine the pulsar luminosity function
numerically: (i) a full dynamical approach; (ii) a snapshot
approach. In the dynamical approach, a simulation is performed in
which a model galaxy of pulsars is seeded according to various
prescriptions of birth locations and initial rotational
parameters. Each of these synthetic pulsars is then ``evolved'' both
kinematically in a model for the Galactic gravitational potential and
rotationally using a model for neutron star spin-down. The resulting
population is then passed through the various detection criteria
\citep[see, e.g.,][]{fk06}. The resulting set of ``detectable'' pulsars
is then compared to the observed sample. The snapshot approach
differs from the dynamical approach in that the pulsars are seeded at
their final positions in the galaxy without assuming anything about
their spin-down or kinematic evolution and thus form a picture of the
present day population. 

The dynamical approach has been used
extensively to study normal pulsars in the galactic disk
\citep{bwhv92, goboh02, fk06, rl10}. These studies considered the
luminosity of pulsars to be described by power-law functions involving
$P$ and $\dot{P}$ with a substantial dispersion to account for
distance uncertainties and beam geometry.  
One of the conclusions by
\citet[][hereafter FK06]{fk06}, also verified by \citet{rl10} is that
the resulting parent population of luminosities appears to be well
described by a log-normal function. As discussed further in Section 3.1,
the log-normal parameters favored by
FK06 (for the base-10 logarithm of the 1400-MHz luminosities) are
a mean of --1.1 and a standard deviation of 0.9. 
One of the goals of the current study is to examine whether these
log-normal parameters are
consistent with the observed populations of GC pulsars.

For pulsars in GCs, where it is difficult to model the effects of
stellar encounters and the cluster potential, the dynamical approach
has so far not been used for radio pulsar population
syntheses. Although such an approach may be tractable in future, we
will adopt in this work a version of the snapshot approach. We will
carry out Monte Carlo simulations that assume all GCs have the same
intrinsic pulsar luminosity function, but a different population
size and use this approach to model the observed sample of
pulsars given the various ranges of luminosities. As we shall see,
this approach provides a remarkably good agreement between the model
and observed luminosity distributions.

Within the snapshot framework, one usually fits the complementary cumulative distribution
function (CCDF) of pulsar luminosity as a power law as $N ( >
L_{1400}) = N_0 L_{1400}^q$, where $L_{1400}$ is the luminosity of the
pulsar at 1400 MHz, $N$ is the number of pulsars having luminosity
value greater than $L_{1400}$, $N_0$ and $q$ are constants. 
We note here that sometimes in literature
\citep{hrskf07, hct10}, the CCDF has been mentioned as the cumulative
distribution function; but actually the cumulative distribution
function (CDF, $N ( \leq L_{1400})$ ) is related to the CCDF as CDF =
$1-$~CCDF. Using this snapshot technique, \citet{hct10} concluded that
the luminosities of MSPs in GCs are different from those in the
galactic disk as the CCDF for GC MSPs is much steeper than that of disk
pulsars. This is a very important conclusion. If correct, it would
imply that the radio luminosity is related to differences in formation
processes between the disk and GC pulsars. The same analysis was
re-performed with more recent distance estimates of GCs and the
resultant CCDF was even steeper \citep{bl10}.

Since GCs are generally at large distances, the luminosity function of
observed pulsars is not as well sampled as in the Galactic disk.  In
the present work we try to account for this incompleteness by
considering GC MSPs as the brighter tail of some intrinsic parent
population. The goal of the current work is to explore the range of
possible distributions that are consistent with the current sample of
GC pulsars. The plan for the rest of this paper is as follows. In
Section \ref{sec:data}, we describe the pulsar sample we use. In
Section \ref{sec:anal}, we present our analysis procedure and its main
results. In Section \ref{sec:constraints}, we investigate additional
constraints on allowed model parameters from observations of diffuse
radio and gamma-ray flux. In Section \ref{sec:comp}, we compare our
results with earlier work. We draw our main conclusions in Section
\ref{sec:conclu}.

\section{The observational sample}
\label{sec:data} 

The luminosity of a pulsar, $L$, can be computed \citep[see, e.g.,][]{lk05} 
from its distance and the mean flux
density $S_{\nu}$ (defined at some observing frequency $\nu$) using the
following geometrical relationship:
\begin{equation}
L = \frac{4 \pi d^2}{\delta} \, {\rm sin^2 }\left(\frac{\rho}{2} \right) \int_{\nu_{\rm low}}^{\nu_{\rm up}} S_{\nu} d \nu.
\end{equation} 
Here $\rho$ is the radius of the (assumed circular)
emission cone,
$\delta$ is the pulse duty cycle ($ = W_{\rm eq}/P$), $P$ is the spin
period of the pulsar and $W_{\rm eq}$ is the equivalent width of the pulse
(i.e. the width of a top-hat shaped pulse having the same area
and peak flux density as the true profile), $\nu_{\rm low}$ and $\nu_{\rm up}$
defines the range of radio frequencies over which the pulsar has been
observed and $d$ is the distance of the pulsar. As it is usually
difficult to determine the values of $\rho$ and $\delta$ reliably, we
define the ``pseudoluminosity''
\begin{equation}\label{eq:pseudo}
L_{\nu} = S_{\nu}~d^2.
\end{equation}
Henceforth, we will use the term luminosity to mean pseudoluminosity.

Among the current sample of 143 GC pulsars, flux density values have
been reported for 107 pulsars. Among these,
three are clearly young isolated objects which more closely resemble
the normal population of pulsars in the Galaxy \citep[for a further 
discussion of this population, see][]{bltmlr11}. We consider here the
sample of 83 pulsars in 10 GCs with spin periods $P \leq 100$~ms and each of these GCs host at least 4 such pulsars. For all 
these objects, the spin and binary properties suggest that the neutron
star has undergone a phase of recycling in the past. 

Among these, for 45 pulsars (14 in 47~Tuc, 4 in M\,3, 5 in M\,5, 5 in
M\,13, 5 in NGC~6752, 3 in NGC~6517 and 9 in M\,28) flux density values have been
measured at 1400 MHz; for 31 pulsars (25 in Terzan 5, 5 in NGC
6440 and 1 in NGC~6517) flux density values have been measured at 1950 MHz and for 7
pulsars (in M\,5) flux density values have been measured at 400 MHz. To
pursue our study of pulsar luminosities at 1400 MHz, we
scale the flux densities measured at other frequencies using the
power-law $S_{\nu} \propto \nu^{\alpha}$,
where $\alpha$ is the spectral index. We then use the model prediction
for $S_{1400}$ as the best estimate of the pulsar's flux density at
this frequency. In these calculations we use the estimated values of $\alpha$ from observed values of fluxes at different frequencies whenever available, otherwise adopt the mean $\alpha$ of GC MSPs (for which flux values
have been reported at multiple frequencies) of $-1.9$. \citet{tbms98}
also obtained mean $\alpha$ of 19 millisecond pulsars to be $-1.9 \pm
0.1$, but their sample contains only two GC pulsars. Once we get
$S_{1400}$, we can calculate $L_{1400}$ if $d$ is known. For GC
pulsars, the distances are taken to be those of their host
clusters. We use the most recent distance estimates from 
the literature.

\begin{table*}\footnotesize
\caption{Fluxes and spectral indices of 107 pulsars in globular
  clusters. $\alpha$ can be calculated for 20 pulsars using the
  central frequency of observations (when ever reported). Mean
  $\alpha$ of pulsars having $P_s \leq $ 100 ms is --1.865 (excluding
  positive $\alpha$ of PSR J1836$-$2354A). We set $\alpha = -1.9$ in the
  present work. From the sample of 107 pulsars, we exclude pulsars
  with spin period $>$ 100 ms; and then exclude pulsars for which the
  host GC contains less than 4 pulsars with $P_s \leq $ 100 ms and known
  flux values. Pulsars which are not used in the present study have
  been written in italics. References are at the end of the table.}
\begin{tabular}{llccccccccc} \hline \hline 
GC & PSR & $P_s$ & $S_{400}$ & $S_{600}$ & $S_{1170}$ &$S_{1400}$ & $S_{1600}$& $S_{1950}$& $\alpha$  \\
& & (ms)  & (mJy) & (mJy) & (mJy) & (mJy) & (mJy) & (mJy) &  \\ \hline 
47 Tuc  &  J0023-7204C    &        5.757      &      1.53(\textit{r1})   &    1.54(\textit{r1})     &     *     &   \textit{0.36}(\textit{r2}) &        *      &      * & -1.352   \\ 
47 Tuc  &  J0024-7204D    &        5.358      &      0.95(\textit{r1})   &    0.55(\textit{r1})     &     *     &   \textit{0.22}(\textit{r2})          &        *      &      * & -1.264  \\ 
47 Tuc  &  J0024-7205E    &        3.536      &      *                   &       *                  &     *     &     0.21(\textit{r2})           &       *      &      * &  * \\ 
47 Tuc  &  J0024-7204F    &        2.624      &      *                   &       *                  &     *     &      0.15(\textit{r2})        &       *      &      * &  * \\ 
47 Tuc  &  J0024-7204G    &        4.040      &      *                   &       *                  &     *     &       0.05(\textit{r2})          &       *      &      * &  * \\ \\ 
47 Tuc  &  J0024-7204H    &        3.210      &      *                   &       *                  &     *     &       0.09(\textit{r2})        &       *      &      * &  * \\ 
47 Tuc  &  J0024-7204I    &        3.485      &      *                   &       *                  &     *     &       0.09(\textit{r2})        &        *     &       * &  * \\ 
47 Tuc  &  J0023-7203J    &        2.101      &      *                   &       *                  &     *     &       0.54(\textit{r2})         &       *      &      * &  * \\ 
47 Tuc  &  J0024-7204L    &        4.346      &      *                   &       *                  &     *     &       0.04(\textit{r2})          &       *     &       * &  * \\ 
47 Tuc  &  J0023-7205M    &        3.677      &      *                   &       *                  &     *     &      0.07(\textit{r2})           &      *     &       * &  * \\ \\ 
47 Tuc  &  J0024-7204N    &        3.054      &      *                   &       *                  &     *     &       0.03(\textit{r2})           &       *      &      * &  * \\ 
47 Tuc  &  J0024-7204O     &       2.643      &      *                   &       *                  &     *     &       0.10(\textit{r2})       &        *     &       * &  * \\ 
47 Tuc  &  J0024-7204Q     &        4.033     &      *                   &       *                  &     *     &        0.05(\textit{r2})         &       *      &      * &  * \\ 
47 Tuc  &  J0024-7203U      &       4.343     &      *                   &       *                  &     *     &        0.06(\textit{r2})         &       *     &       *   &  *  \\         
\textit{NGC 1851} & \textit{J0514-4002}       &       4.990     &      0.28(\textit{r3})   &       *                  &     *     &     *   & *  &  0.0056(\textit{r3})  &  -2.568 \\  \\              
\textit{M 53}   &   \textit{B1310+18}         &       33.163    &      1.0(\textit{r4})    &       *                  &     *     &     *              &      *     &       *  &  *  \\                
M 3    &   J1342+2822A      &       2.545     &      *                   &       *                  &     *     &     0.007(\textit{r5})   &      *     &       *  &  * \\                
M 3    &   J1342+2822B      &       2.389     &      *                   &       *                  &     *     &    0.014(\textit{r5})    &     *     &       *  &   * \\                
M 3    &   J1342+2822C      &       2.166     &      *                   &       *                  &     *      &    0.006(\textit{r5})    &     *     &       *   &  * \\               
M 3   &    J1342+2822D      &       5.443     &      *                   &       *                  &     *      &     0.010(\textit{r5})    &     *     &       *   &  * \\   \\           
M 5   &    B1516+02A        &       5.554     &      0.5(\textit{r6})    &       *                  &   0.155(\textit{r7})   &     0.120(\textit{r5})    &     *     &       *   & -1.161   \\        
M 5   &    B1516+02B        &       7.947     &      0.3(\textit{r6})    &       *                  &   0.027(\textit{r7})   &     0.025(\textit{r5})    &     *      &      *   &  -2.132 \\         
M 5  &     J1518+0204C      &       2.484     &      *                   &       *                  &     *       &     0.039(\textit{r5})    &     *     &       *   & *  \\         
M 5   &    J1518+0204D      &       2.988     &      *                   &       *                  &     *      &      0.008(\textit{r5})    &     *     &       *   &  * \\               
M 5   &    J1518+0204E      &       3.182     &      *                   &       *                  &     *      &      0.010(\textit{r5})    &     *     &       *   &  * \\  \\              
\textit{M 4}   &    \textit{B1620-26}         &       11.076    &     15(\textit{r8})      &       7.2(\textit{r9})   &     *      &      1.6(\textit{r10})      &     *     &       *   & -1.744  \\ 
M 13  &    B1639+36A        &      10.378     &     3.0(\textit{r4})      &      *     &      *      &      0.140(\textit{r5})    &     *     &       *   & -2.486  \\                         
M 13  &    B1639+36B        &      3.528      &     *        &      *     &      *     &       0.022(\textit{r5})    &     *     &  *     &     *  \\               
M 13  &    J1641+3627C      &      3.722      &     *        &      *     &      *     &       0.030(\textit{r5})    &     *     &       *   &  *  \\               
M 13  &    J1641+3627D      &      3.118      &     *        &      *     &      *     &       0.024(\textit{r5})    &     *     &       *   &  *  \\   \\             
M 13  &    J1641+3627E      &      2.487      &     *        &      *     &      *     &       0.010(\textit{r5})    &     *     &       *   & *   \\       
\textit{M 62}  &    \textit{J1701-3006A}      &      5.242      &     *        &      *     &      *     &       0.4(\textit{r11})           &     *     &       *    &  *  \\               
\textit{M 62}  &    \textit{J1701-3006B}      &      3.594      &     *        &      *     &      *     &       0.3(\textit{r11})      &     *     &       *      & *   \\          
\textit{M 62}  &    \textit{J1701-3006C}      &      7.613      &     *       &       *     &      *     &       0.3(\textit{r11})     &      *     &       *       &  *  \\           
\textit{NGC 6342} & \textit{B1718-19}   &  1004.04   &  0.253(\textit{r12})   &   0.550(\textit{r12}) &      *    &   0.278(\textit{r12})     &      0.18(\textit{r12})     &      *    & -0.338 \\   \\     
\textit{NGC 6397} & \textit{J1740-5340}      &       3.650      &       *      &        *     &      *     &       1.0(\textit{r13})         &      *     &       *      &  *  \\ 
Ter 5  &  J1748-2446A      &      11.563      &      *       &       5(\textit{r14})      &     *      &      0.61(\textit{r15})     &     *     &       1.020(\textit{r16})  &  -1.572  \\           
Ter 5  &  J1748-2446C      &      8.436       &      *       &       *      &     *      &      *                  &     *     &       0.360(\textit{r16}) &  * \\            
Ter 5  &  J1748-2446D     &       4.714       &      *       &       *      &     *      &      *                 &     *     &       0.041  (\textit{r16}) &  *  \\         
Ter 5  &  J1748-2446E     &       2.198       &      *      &        *      &     *      &      *             &     *     &       0.048(\textit{r16})   &  *  \\  \\         
Ter 5  &  J1748-2446F     &       5.540       &      *       &       *     &      *      &      *           &     *     &       0.035(\textit{r16})   & *   \\         
Ter 5  &  J1748-2446G     &       21.672      &      *       &       *      &     *      &      *        &     *     &       0.015(\textit{r16})   &  *  \\         
Ter 5  &  J1748-2446H     &       4.926       &      *       &       *     &      *      &      *         &     *     &       0.015(\textit{r16})   &  *  \\          
Ter 5  &  J1748-2446I     &       9.570       &      *       &       *     &      *       &     *          &     *     &       0.029(\textit{r16})    & *   \\         
Ter 5 &   J1748-2446J     &       80.338      &      *       &       *     &      *      &      *        &     *     &       0.019(\textit{r16})     & *  \\      \\   
Ter 5  &  J1748-2446K     &       2.970       &      *       &       *     &      *      &      *           &     *     &       0.040(\textit{r16})    &  *  \\      
Ter 5 &   J1748-2446L     &       2.245       &      *       &       *     &      *      &      *         &     *     &       0.041(\textit{r16})    &  *   \\          
Ter 5  &  J1748-2446M     &       3.570       &      *       &       *     &      *      &      *        &     *     &       0.033(\textit{r16})    & *   \\          
Ter 5  &  J1748-2446N     &       8.667       &      *       &       *     &      *      &      *         &     *     &       0.055(\textit{r16})     &  *  \\         
Ter 5  &  J1748-2446O     &       1.677       &      *       &       *     &      *      &      *        &     *     &       0.120(\textit{r16})    &  *  \\     \hline

\end{tabular}
\label{tb:pulsar_props}
\end{table*}

\addtocounter{table}{-1}
\begin{table*}\footnotesize
\caption{(continued).}
\begin{tabular}{llccccccccc} \hline \hline 
GC & PSR & $P_s$ & $S_{400}$ & $S_{600}$ & $S_{1170}$ &$S_{1400}$ & $S_{1600}$& $S_{1950}$& $\alpha$  \\
& & (ms)  & (mJy) & (mJy) & (mJy) & (mJy) & (mJy) & (mJy) &  \\ \hline     
Ter 5  &  J1748-2446P     &       1.729       &      *      &        *     &      *     &       *       &      *     &       0.077(\textit{r16})    &  *  \\           
Ter 5 &   J1748-2446Q     &       2.812       &      *      &        *     &      *     &       *       &      *     &       0.027(\textit{r16})    & *   \\          
Ter 5 &   J1748-2446R     &       5.028       &      *      &        *     &      *     &       *      &      *     &       0.012(\textit{r16})    & *   \\       
Ter 5  &  J1748-2446S     &       6.117       &      *      &        *     &      *     &       *        &      *     &       0.018(\textit{r16})    &  *  \\          
Ter 5  &  J1748-2446T     &       7.085       &      *      &        *     &      *     &       *         &      *     &       0.020(\textit{r16})    & *  \\      \\     
Ter 5  &  J1748-2446U     &       3.289       &      *      &        *     &      *     &       *      &       *     &       0.016(\textit{r16})    &  *  \\        
Ter 5 &   J1748-2446V     &       2.072       &      *      &        *     &      *     &       *        &       *     &       0.071(\textit{r16})   &  *  \\         
Ter 5  &  J1748-2446W     &       4.205      &      *      &        *     &      *     &       *       &     *    & 0.022(\textit{r16})     &   * \\    
Ter 5  &  J1748-2446X     &       2.999      &      *      &        *     &      *     &       *       &       *     &       0.018(\textit{r16})    &  *  \\       
Ter 5  &  J1748-2446Y     &       2.048      &      *      &        *     &      *     &       *      &       *     &       0.016(\textit{r16})    &  *  \\    \\           
Ter 5  &  J1748-2446ad    &       1.396      &      *      &        *     &      *     &       *     &       *     &       0.08(\textit{r17})        &  *  \\ 
\textit{NGC 6440} & \textit{B1745-20}    &   288.603   &   10(\textit{r18})    &    *    &     *    &   0.37(\textit{r15})    &    1.5(\textit{r18})    &       0.37(\textit{r19})  & -1.920  \\  
NGC 6440 & J1748-2021B    &        16.760       &       *      &        *    &       *     &       *         &        *     &       0.047(\textit{r19})     &  *  \\  
NGC 6440 & J1748-2021C    &        6.227            &       *      &        *    &       *     &       *          &        *     &       0.044(\textit{r19})     & *   \\  
NGC 6440 & J1748-2021D    &        13.496       &       *      &        *    &       *     &       *         &        *     &       0.075(\textit{r19})     &  *  \\  \\ 
NGC 6440 & J1748-2021E     &       16.264            &       *      &        *    &       *     &       *         &        *     &       0.023(\textit{r19})     & *   \\  
NGC 6440 & J1748-2021F    &        3.794       &       *      &        *    &      *      &      *           &     *     &       0.017(\textit{r19})     &  *  \\  
\textit{NGC 6441} & \textit{J1750-37A}      &        111.608       &       *      &        *    &       *     &       *          &       *     &       0.059(\textit{r19})     &  *  \\  
\textit{NGC 6441} & \textit{J1750-3703B}    &        6.074      &       *      &        *    &       *     &       *         &       *     &       0.037(\textit{r19})     &  *  \\  
\textit{NGC 6441} & \textit{J1750-3703C}    &        26.569          &       *      &        *    &       *     &       *         &       *     &       0.015(\textit{r19})      &  *  \\  \\ 
\textit{NGC 6441} & \textit{J1750-3703D}    &        5.140          &        *      &        *     &      *     &       *       &       *     &       0.010(\textit{r19})     &  *  \\  
NGC 6517 & J1801-0857A    &        7.176             &        *      &        *     &      *     &       0.036(\textit{r20})        &       *     &       0.020(\textit{r20})        &   -1.648 \\          
NGC 6517 & J1801-0857B    &        28.961            &        *      &        *     &      *     &       0.012(\textit{r20})        &       *     &       0.009(\textit{r20})        &  -0.806  \\          
NGC 6517 &  J1801-0857C   &         3.739          &         *     &         *    &       *    &        0.012(\textit{r20})       &        *     &       0.007(\textit{r20})       &   -1.511 \\  
NGC 6517 &  J1801-0857D   &         4.226         &         *     &         *    &       *    &        *        &        *     &       0.011(\textit{r20})        &   * \\   \\ 
\textit{NGC 6539} & \textit{B1802-07}       &     23.101      &     3.1(\textit{r21})    &    1.0(\textit{r21})   &     *     &    0.6 (\textit{r21})    &       *     &       *        &  -1.213  \\       
\textit{NGC 6544} & \textit{J1807-2459A}    &    3.059      &        *      &        *     &      *     &       1.3(\textit{r22})       &      *     &       *        & * \\             
\textit{NGC 6624} & \textit{B1820-30A}    &    5.440   &    16(\textit{r9})     &   6.8(\textit{r9})   &    *     &    0.72(\textit{r9})    &      0.31(\textit{r23})   &      *      & -2.922  \\            
\textit{NGC 6624} & \textit{B1820-30B}    &  378.596   &    2.2(\textit{r21})    &   1.0(\textit{r21})   &      *     &     0.07(\textit{r23})    &      0.07(\textit{r23})   &      *      & -2.654  \\   
M 28   &   B1821-24A      &        3.054       &        30(\textit{r24})     &        *     &      *     &       0.94(\textit{r25})    &      *     &       *      & -2.764  \\    \\         
M 28   &   J1824-2452B     &       6.547            &        *      &        *     &      *     &       0.07(\textit{r25})    &      *     &       *     &  *  \\             
M 28   &   J1824-2452C    &        4.159         &        *      &        *     &      *     &       0.17(\textit{r25})    &      *     &       *     &  *  \\              
M 28   &   J1824-2452D    &        79.832       &        *      &        *     &      *     &       0.05(\textit{r25})    &      *      &      *     & *   \\            
M 28   &   J1824-2452E    &        5.420             &        *      &        *     &      *     &       0.06(\textit{r25})    &      *      &      *     &  *  \\             
M 28   &   J1824-2452F    &        2.451           &        *      &        *     &      *     &       0.08(\textit{r25})    &      *     &       *     &  *  \\  \\             
M 28  &   J1824-2452G     &       5.909           &      *       &       *      &     *      &       0.05(\textit{r25})    &      *     &       *     &  *  \\            
M 28  &    J1824-2452H    &        4.629          &        *      &        *     &      *      &      0.06(\textit{r25})    &      *     &       *     & *   \\            
M 28  &    J1824-2452J    &        4.039         &       *      &        *     &      *      &      0.07(\textit{r25})    &      *     &       *     &  * \\  
\textit{M 22}   &   \textit{J1836-2354A}    &    3.354   &     *    &     *     &      *      &      0.040(\textit{r20})           &      *     &       0.043(\textit{r20})     &  0.203  \\             
\textit{M 22}   &   \textit{J1836-2354B}    &    3.232   &     *   &     *    &      *     &      0.200(\textit{r20})          &      *     &       0.073(\textit{r20})       & -2.826  \\   \\               
NGC 6749 & J1905+0154A    &        3.193        &       *     &       *     &      *      &     0.023(\textit{r5})   &     *     &     *      & * \\              
NGC 6749 & J1905+0154B    &        4.968             &       *      &        *     &      *      &      0.006(\textit{r5})   &      *     &       *     &  *  \\             
NGC 6752 & J1911-5958A    &        3.266       &       *      &        *     &      *      &      0.21(\textit{r26})    &      *     &       *     &  *  \\              
NGC 6752 & J1910-5959B    &        8.358           &       *      &        *     &      *      &      0.05(\textit{r26})    &      *     &       *     &  *  \\              
NGC 6752 & J1911-6000C    &        5.277      &        *      &        *     &      *      &      0.24(\textit{r26})    &      *      &      *     &  *  \\    \\            
NGC 6752 & J1910-5959D    &        9.035     &        *      &        *     &      *      &      0.05(\textit{r26})    &      *     &       *     &  *  \\               
NGC 6752 & J1910-5959E    &        4.572       &        *      &        *     &      *      &      0.07(\textit{r26})    &      *     &       *    &  *  \\                          
\textit{M 71}  &    \textit{J1953+1846A}    &    4.888         &       *      &        *     &      *      &      0.059(\textit{r5})    &      *     &       *    &  * \\              
\textit{M 15}  &    \textit{B2127+11A}      &   110.665       &       1.7(\textit{r27})     &       *     &      *      &      0.2(\textit{r28})     &      *     &       *    &  -1.797  \\      
M 15  &    B2127+11B      &        56.133        &       1.0(\textit{r27})    &        *     &      *      &      *              &      *     &       *    &  *  \\           \hline 

\end{tabular}
\end{table*}

\addtocounter{table}{-1}
\begin{table*}\footnotesize
\caption{(continued).}
\begin{tabular}{llccccccccc} \hline \hline 
GC & PSR & $P_s$ & $S_{400}$ & $S_{600}$ & $S_{1170}$ &$S_{1400}$ & $S_{1600}$& $S_{1950}$& $\alpha$  \\
& & (ms)  & (mJy) & (mJy) & (mJy) & (mJy) & (mJy) & (mJy) &  \\ \hline 
M 15  &    B2127+11C      &        30.529        &       0.64(\textit{r27})     &       *     &      *      &      *             &      *     &       *    &  *  \\            
M 15   &   B2127+11D      &        4.803         &       0.34(\textit{r27})     &       *     &      *      &      *             &      *     &       *    &  *  \\             
M 15   &   B2127+11E      &        4.651        &        0.24(\textit{r27})     &       *     &      *      &      *            &      *     &       *    &  *  \\            
M 15   &   B2127+11F      &        4.027        &        0.14(\textit{r27})     &       *     &      *      &      *            &      *     &       *    &  *  \\              
M 15   &   B2127+11G      &        37.660        &        0.13(\textit{r27})     &       *     &      *      &      *            &      *     &       *    &  *  \\     \\  
M 15   &   B2127+11H      &        6.743         &        0.16(\textit{r27})     &       *     &      *      &      *              &     *     &       *    &  *  \\   
\textit{M 30}   &   \textit{J2140-2310A}    &        11.019      &        *       &       *     &      *      &      0.08(\textit{r29})            &      *     &       *    &  *  \\   \hline \hline

\end{tabular}
\vskip 0.3cm
{\footnotesize{References : (1) \textit{r1} : \citet{rlm95}, (2) \textit{r2} : \citet{clf00}, (3) \textit{r3} : \citet{frg07}, (4) \textit{r4} : \citet{kapw91}, (5) \textit{r5} : \citet{hrskf07}, (6) \textit{r6} : \citet{awkp97}, (7) \textit{r7} : \citet{fwvh08}, (8) \textit{r8} : unpublished (\textit{http://www.atnf.csiro.au/research/pulsar/psrcat/expert.html}), (9) \textit{r9} : \citet{tbms98}, (10) \textit{r10} : \citet{kxl98}, (11) \textit{r11} : \citet{pam03}, (12) \textit{r12} : Averaged over variations with the orbital phase \citep{lbhm93}, (13) \textit{r13} : \citet{alm01}, (14) \textit{r14} : \citet{ljmsa90}, (15) \textit{r15} : \citet{hfs04}, (16) \textit{r16} : \citet{rhs05},  (17) \textit{r17} : \citet{hrs06}, (18) \textit{r18} : \citet{lma96}, (19) \textit{r19} : \citet{frb08}, (20) \textit{r20} : \citet{lrfs11}, (21) \textit{r21} : \citet{lylg95}, (22) \textit{r22} : \citet{rghm01},
(23) \textit{r23} : \citet{bblgf94}, (24) \textit{r24} : \citet{ffb91}, (25) \textit{r25} : \citet{bgn06}, (26) \textit{r26} : \citet{cpl06}, (27) \textit{r27} : \citet{and93}, (28) \textit{r28} : \citet{wkm89}, (29) \textit{r29} : \citet{rsb04}     }}
\end{table*}

\begin{figure}
\centerline{\psfig{figure=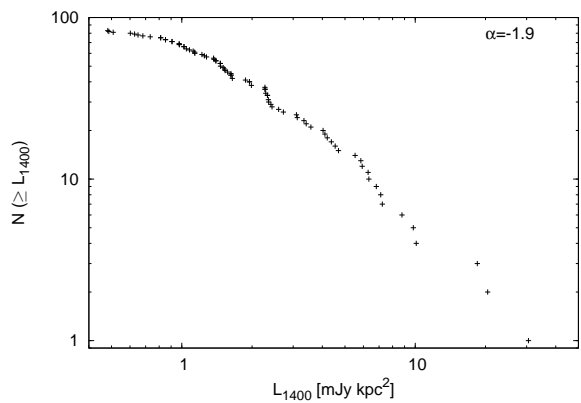,width=8cm,angle=0}}
\caption{Complementary cumulative distribution of 1400-MHz luminosities
for the sample of globular cluster pulsars. Flux density values measured at other
frequencies have been converted to $S_{1400}$ using 
$S_\nu \propto \nu^{\alpha}$ for $\alpha = -1.9$. We have checked that a change in the value of $\alpha \sim \pm 0.3$ does not make any visible change in the shape of the plot.} \label{fig:obscdf_allalpha}
\end{figure}

Our complete list of GC pulsar flux and spectral
parameters used in this section, and the remainder of the paper
is given in Table \ref{tb:pulsar_props}. While compiling this list,
we confirmed the earlier conclusions by \citet{hrskf07} that
the choice of $\alpha$ in a realistic range does not affect the
complementary cumulative distribution (CCD) of luminosities
significantly. Using 37 isolated GC pulsars, \citet{hrskf07} found
that an arbitrary choice of $\alpha$ in the range of --1.6 to --2.0
does not affect the shape of the CCD. We arrive at the same conclusion
with our sample using different values of $\alpha$ as --1.6, --1.9 and
--2.2. We perform Kolmogorov-Smirnov (KS) tests between the luminosity distributions
obtained with different choice of $\alpha$. The KS test returns 
a statistic $P_{\rm KS}$ which gives the probability that the two samples
are drawn from the same distribution \citep[for details, see][]{press07}.
In this case, $P_{\rm KS}$ is always greater than $0.997$ when we compare any two
luminosity distributions from the three obtained with $\alpha = -1.6$,
$-1.9$ and $-2.0$; as an example, the distribution obtained with $\alpha = -1.9$ is shown in 
Fig.~\ref{fig:obscdf_allalpha}.

The absence of any significant difference in the luminosity
distribution for a realistic range of $\alpha$ also supports our
choice of $\alpha = -1.9$ for this work, which is not too
different than the favored choice ($\alpha = -1.8$) of other studies
\citep{hct10, mkkw00}.  As we are studying luminosities only at 1400
MHz, henceforth, we shall denote $L_{1400}$ simply by $L$. 

  Note also that, throughout this paper, we will be concerned
  with the possible forms of the luminosity distribution of cluster
  pulsars. Correlations between luminosity and other pulsar parameters
  are not discussed in any detail here. The reason for this is that,
  as for the pulsar population in the Galaxy correlations in the
  observed pulsar samples are not apparent due to the presence of
  distance errors and beaming uncertainties (see Lorimer et al.~1993
  \nocite{lbdh93} for a discussion). We did not find any correlation
  between luminosity and spin period of the 83 recycled pulsars used
  in the present work. As mentioned earlier, $\dot{P}$ measurements
  for globular cluster pulsars are affected by cluster potential, so
  can not easily be used to study intrinsic properties of the
  pulsars. For the remainder of this paper, we proceed with the
  underlying assumption that there exists a single luminosity function
  for all globular cluster pulsars, and attempt to explain the
  observed luminosities in this way. As we will demonstrate, the
  data are remarkably consistent with this simple idea. However, the
  wide ranges of possible model parameters that are consistent with
  the data do not rule out the idea that the parent luminosity function
  may vary from cluster to cluster.

\section{Analysis} 
\label{sec:anal}

With the data described above, we aim to find luminosity distribution
functions whose brighter tail can be considered as the observed
luminosity distribution of GC pulsars, assuming that the parent
luminosity distribution is the same for all GCs. To do so, for
each GC, we first generate a synthetic sample of $N_{\rm trial, i}$
pulsar luminosities from a chosen distribution function until we get
$C \times N_{\rm obs, i}$ pulsars having simulated luminosities
greater than the observed minimum luminosity for that GC. This
multiplication by the constant `$C$' (100--1000) is done to minimize
statistical variations. In this notation, $i$ is the GC index 
$N_{\rm obs, i}$ is the observed number of pulsars in the GC that we
consider,
\begin{equation}
L_{\rm sim,tot} = \frac{1}{C} \displaystyle\sum\limits_{j=1}^{N_{\rm trial,i}} L_{\rm sim, j}
\label{eq:lum_pred}
\end{equation} 
is the total luminosity and
\begin{equation}
S_{\rm sim, tot} = \frac{1}{C} \displaystyle\sum\limits_{j=1}^{N_{\rm trial, i}} S_{\rm sim, j}
\label{eq:flux_pred}
\end{equation} 
is the total flux in the $i^{\rm th}$ GC. Here $L_{\rm sim, j}$ and
$S_{\rm sim, j}$ are the simulated luminosities and corresponding
fluxes. After we perform the simulation for all 10 GCs, we compare the
simulated luminosities with the observed luminosities of 83 pusars by
performing KS and $\chi^2$ tests. As mentioned earlier, the KS test
can be used to test the hypothesis that two distributions differ, with
a low value of KS probability $P_{\rm KS}$ suggesting a mismatch. The
$\chi^2$ statistic uses binned data and
compares the values of the two distributions at
each bin; here a low value of $\chi^2$ implies a good agreement. Here we
divide the luminosity range 0.1--1000 ${\rm mJy ~kpc^2}$ into 36
logarithmically equispaced bins. $N_{\rm trial, i}/C$ is the predicted
number of total pulsars in that GC which we call as $N_{\rm rad,
  i}$. 

A key assumption in our present analysis is that each GC has been
searched down to the level of the faintest observable pulsar in that
particular cluster. This assumption provides a good approximation to
the actual survey sensitivity in each cluster, and was made primarily
due to the lack of currently published detail of several of the
globular cluster surveys so far. The assumption greatly simplifies our
modeling procedure, since it means that we do not have to consider
variations in sensitivity due to other factors (for example
scintillation, eclipsing binary systems etc.). This simple approach is
appropriate for the purposes of the current work where we are simply
trying to assess the range of luminosity functions compatible with the
data. A more rigorous study which takes account of the survey
thresholds in detail may well be able to narrow the range of possible
model parameters found here, and should certainly be carried out when
more details of the surveys are published, but is beyond the scope of
the current work.

\subsection{Log-normal luminosity function}

\begin{figure}
\centerline{\psfig{figure=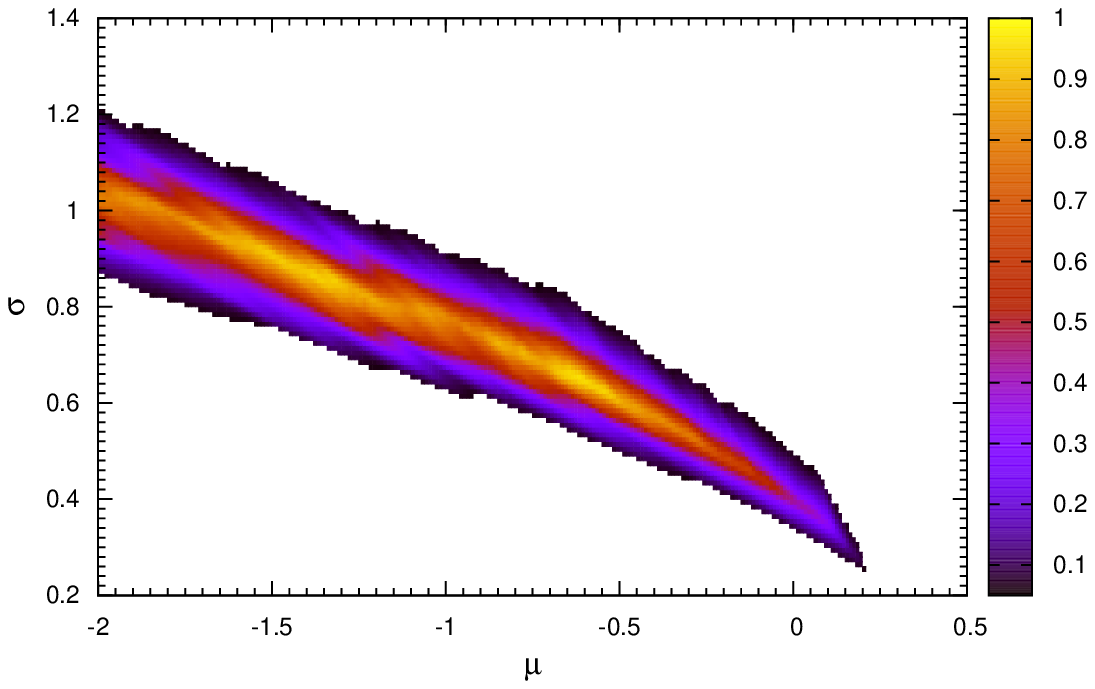,width=7cm,angle=0}}
\centerline{\psfig{figure=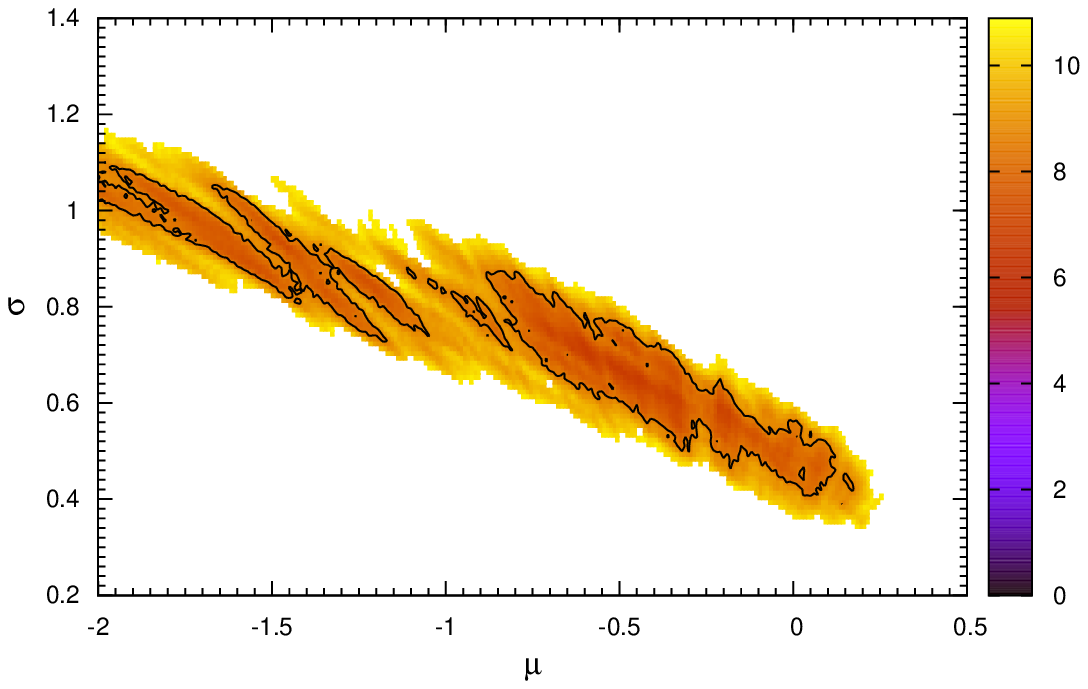,width=7cm,angle=0}}
\caption{Top: variation of $P_{\rm KS}$ (for $P_{\rm KS} \geq 0.05$)
  with $\mu$ and $\sigma$. Bottom: variation of $\chi^{2}$ (within $2
  \sigma$ about the minimum value of $\chi^{2}$) with $\mu$ and
  $\sigma$, the parameters of the log-normal distribution. A $1
  \sigma$ contour is also shown.} \label{fig:kschisq_lognorm}
\end{figure}

We begin by testing a log-normal luminosity function, where the
probability density function (PDF) 
\begin{equation}
f_{\rm log-normal} \, (L)  =  \frac{\log_{10}e}{L} \, \frac{1}{\sqrt{2 \pi \sigma^2}} \, \exp\left[{\frac{-(\log_{10} L - \mu)^2}{2 \sigma^2}}\right],
\label{eq:lognorm_dist_def}
\end{equation}
where, as usual, $\mu$ is the mean of the distribution and $\sigma$
is the standard deviation. For this choice of distribution,
we find that $C=100$ is sufficient to minimize statistical
fluctuations.  The variation of $P_{\rm KS}$ and $\chi^{2}$ with $\mu$
and $\sigma$ are shown in Fig.~\ref{fig:kschisq_lognorm}. It is clear
that there is a wide range of values of $\mu$, $\sigma$ for which the
simulated luminosity distributions agree well with the observed
sample. For the two statistical tests, good agreement is given when
$P_{\rm KS}$ has high values and $\chi^{2}$ is small. As expected, the
region of $\mu-\sigma$ parameter space encompassed by $P_{\rm
  KS}>0.05$ is essentially the same as the contours encompassing the
95\% probability values around the $\chi^2$ minimum.

For this distribution, and for the purposes of later discussion, we
define three models based on particular parameter choices.  Model 1
uses the parameters found by FK06 ($\mu=-1.1$ and $\sigma=0.9$) from
which we find $P_{\rm KS} = 0.15 $ and $\chi^2 = 9.4 $. Model 2, for
which $\mu = -0.61$ and $\sigma = 0.65$ returns the maximum value of $P_{\rm KS} = 0.98$
with a $\chi^2 = 7.9$. Model 3, for which $\mu = -0.52$ and $\sigma =
0.62$, returns a minimum value of $\chi^2 = 6.3$ and has $P_{\rm KS}
= 0.37$.

\begin{figure}
\centerline{\psfig{figure=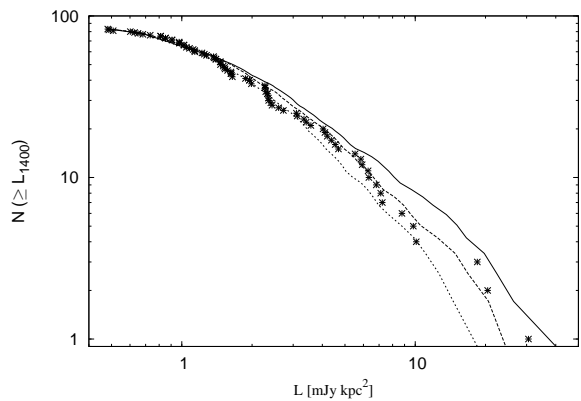,width=8cm,angle=0}}
\caption{Observed luminosity distribution with simulated luminosity distributions generated with log-normal distribution function for three different sets of $\mu$ and $\sigma$ which are defined as models 1 (upper curve), 2 (lower curve) and 3 (middle curve); see text for details. ``$\ast$''s represent the observed distribution.} 
\label{fig:obssim_cdflognormal}
\end{figure}

In Fig.~\ref{fig:obssim_cdflognormal} we compare these three models with the observed data. As expected, all models match
well. While model 3 provides the closest match by eye, the statistical
results mentioned above do not rule out either model 1 or model 2.
The FK06 luminosity model parameters (model 1), therefore, are consistent
with the observed CCD.

\subsection{Power-law luminosity function}

As mentioned earlier, power-law luminosity functions have been used
by a number of authors. It is therefore of great interest to see how
the power-law compares to log-normal for the GC pulsars.
The PDF of the power law distribution
\begin{equation} 
f_{\rm power-law} \, (L)  = \frac{\beta L_{\rm min}^{\beta}}{L^{\beta + 1}},
\label{eq:singlepow_dist_def}
\end{equation}
where $L_{\rm min}$ is the minimum value of $L$ and $\beta$ the
power-law index. This abrupt cut-off required to avoid divergence
when integrating this function over all $L$ is somewhat unphysical,
but nevertheless can be used to parameterize the luminosities 
in an independent way to the log-normal. 
We perform simulations over a range of $L_{\rm min}$
0.003---0.48 ${\rm mJy~kpc^2}$, as 0.48 ${\rm mJy~kpc^2}$ is the observed
minimum luminosity among GC pulsars in our sample, and the lower value
of $L_{\rm min}$ is chosen somewhat arbitrarily. 
For this model, we found that $C=1000$ is required
to minimize statistical fluctuations. Unlike the log-normal
model, we found that the power law distribution occasionally
produced pulsars with large luminosities $L \gg 100$~mJy~kpc$^2$
which biased some of our preliminary simulation runs. To avoid this difficulty,
we imposed a maximum luminosity of 50~mJy~kpc$^2$. No GC pulsar
is currently known with $L>20$~mJy~kpc$^2$, and our results are insensitive to the exact choice of the maximum luminosity cutoff
over the range 20--500~mJy~kpc$^2$.

\begin{figure}
\centerline{\psfig{figure=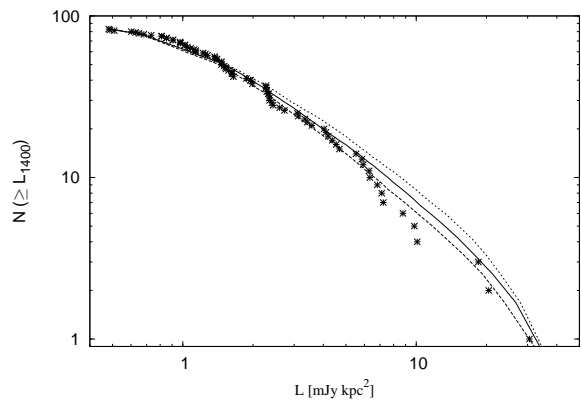,width=8cm,angle=0}}
\caption{Observed luminosity distribution with simulated luminosity distributions generated with a power-law distribution function for three different set of $L_{\rm min}$ and $\beta$ which are defined as models 4 (middle curve), 5 (lower curve) and 6 (upper curve); see text for details. ``$\ast$"s represent the observed distribution. The mismatch between the simulated and observed CCDs appear only when the number of pulsars are $\lesssim 10$.} \label{fig:obssim_cdfpowlawb}
\end{figure}

The nominal best parameter values give $P_{\rm KS}= 0.81$ ($\chi^2 = 8.7$) for $\beta =
0.92$, $L_{\rm min} = 0.017 {\rm~ mJy~kpc^2}$ (model 4) and minimum
$\chi^2 = 8.0$ ($P_{\rm KS}= 0.56$) for $\beta = 1.01$, $L_{\rm min} =
0.022{\rm~ mJy~kpc^2}$ (model 5). Our values of $\beta$ (which give
good fits) are not too different from the conventional values $\beta+1
= 2$. For example, the best fit value of the analysis of \citet{fg00}
for Terzan~5 pulsars was $\beta+1 = 1.85$. We have seen that for $\beta \sim 1$, the
fit does not depend much on $L_{\rm min}$ which again agrees with the
results of \citet{fg00}. As an additional point of reference, we also
consider the nominal power-law parameters discussed by \citet{fg00},
i.e.~$\beta = 0.85$ and $L_{\rm min} = 0.03{\rm~ mJy~kpc^2}$ which we
refer to as model 6. For this pairing, $P_{\rm KS} = 0.43$ and $\chi^2
= 9.0$. We need to remember here \citet{fg00} did not put any
constraint on the maximum value of the luminosity.  In
Fig.~\ref{fig:obssim_cdfpowlawb} we compare the observed CCD with
simulated CCDs for models 4-6. Statistically,
the agreement between simulated and observed distributions is almost
as good as that for log-normal distributions.

%%%%%%%%%%%%%%%%%%%%%%%%%%%%%%%%%%%%%%

\subsection{Exponential Distribution}

\begin{figure}
\centerline{\psfig{figure=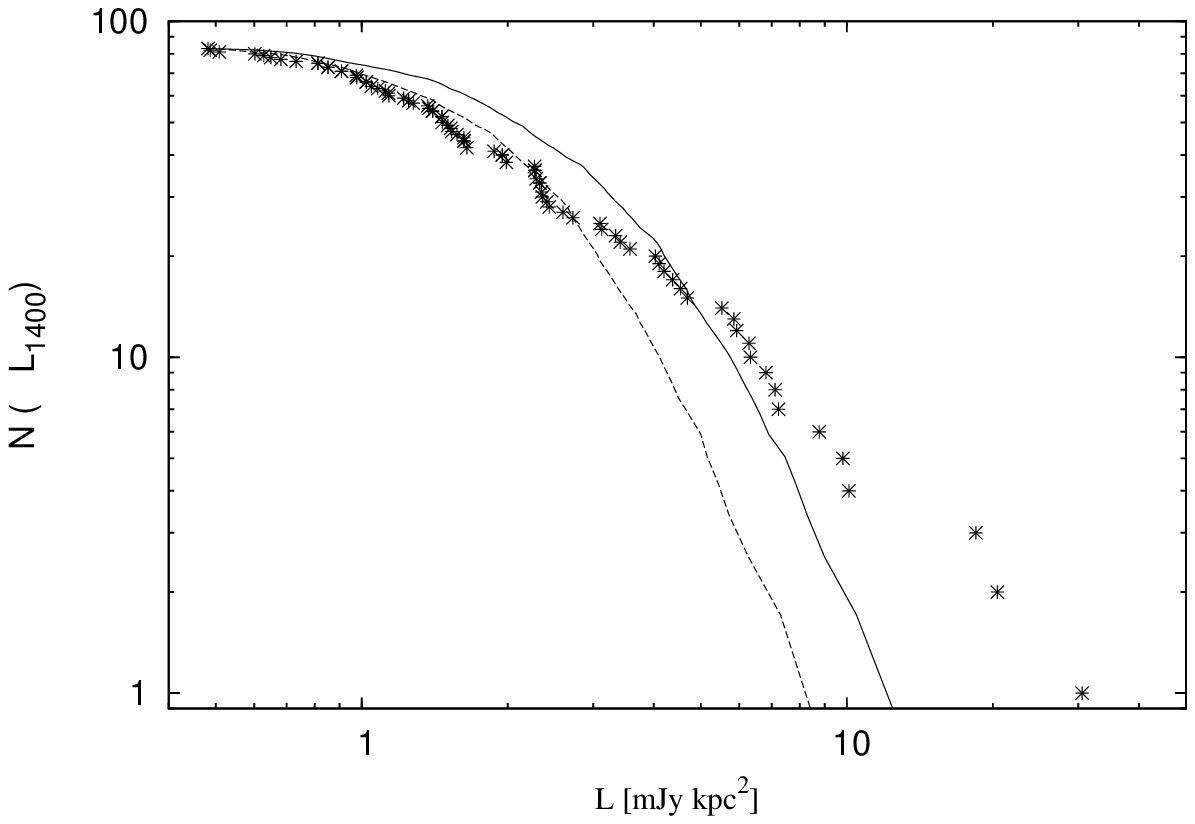,width=8cm,angle=0}}
\caption{Observed luminosity distribution with simulated luminosity distributions generated with exponential distribution function with $\lambda = 0.676 {\rm ~mJy^{-1}~kpc^{-2}}$ (lower curve) and $\lambda = 0.439 {\rm~ mJy^{-1}~kpc^{-2}}$ (upper curve). ``$\ast$"s represent the observed distribution.} \label{fig:obssim_cdfexp}
\end{figure}

The above models characterized the luminosity function in terms
of two parameters. For completeness, we also consider a simple
one-parameter model, the exponential distribution with PDF
\begin{equation}
f_{\rm exponential} \, (L_{1400})  =  \lambda e^{-\lambda L}.
\label{eq:exp_dist_def}
\end{equation} 
Here $1/{\lambda}$ is the mean of the distribution, and we
find that $C=100$ is enough to get rid of statistical fluctuations.
For this model, the maximum value of $P_{\rm KS} = 0.17$ is obtained
for $\lambda = 0.676 {\rm~ mJy^{-1}~kpc^{-2}}$ with a corresponding $\chi^2 = 13.7$. The
minimum value of $\chi^2 = 11.94$ is found for $\lambda =
0.439 {\rm~ mJy^{-1}~kpc^{-2}}$ with a $P_{\rm KS} = 0.00028$. In Fig.~\ref{fig:obssim_cdfexp} we compare
the observed CCD with the simulated CCDs with these two values of
$\lambda$. It is clear that the simulated distribution never matches
with the observed one very well. We therefore do not consider the exponential
distribution further in this work and focus the remainder of the discussion
on the log-normal and power-law distributions.

\section{Model predictions and constraints}
\label{sec:constraints} 

In the previous section, we found that there is a large family of
possible luminosity parameters that are consistent with the observed
distribution of GC pulsar luminosities. These ranges translate to a
variety of different predictions for the population sizes in each
GC. This can be seen from a comparison of the predicted parameters for
each GC using the log-normal parameter choices (models 1, 2 and 3) in
Table \ref{tb:predlnorm} and the power-law parameter combinations
(models 4, 5 and 6) in Table \ref{tb:predplawLmax}.  
In this section, we
try to place further constraints on these parameters by examining the
predictions for the diffuse radio and gamma-ray fluxes separately.

%%%%%%%%%%%

\begin{table*}\footnotesize
\caption{Population estimations and predictions using log-normal luminosity
functions. For each cluster, we list the predicted number of potentially
observable radio pulsars ($N_{\rm rad}$), the predicted total diffuse
radio flux ($S_{\rm sim,tot}$), the predicted gamma-ray luminosity
($L_{\rm \gamma,sim}$) for three different choices of efficiency
($\eta_\gamma$). See text for further details. To compute uncertainties
in $N_{\rm rad}$, we assume that they are dominated by the 
statistical noise in the observed number of pulsars, $N_{\rm obs}$.
The uncertainty in $N_{\rm rad}$ is then simply
$N_{\rm rad}/\sqrt{N_{\rm obs}}$. Also listed for each model is
$N_{10}$, the total population estimate for these 10 GCs.}
\begin{tabular}{|l|c|c|c|c|c|} \hline \hline 
Cluster  & $N_{\rm rad}$  & $S_{\rm sim, tot}$  &  \multicolumn{3}{c}{$L_{\rm \gamma,sim}$}   \\  
  &  &   & $ \langle \eta_{\gamma} \rangle = 0.08 $  & $ \langle \eta_{\gamma} \rangle = 0.06 $ &  $ \langle \eta_{\gamma} \rangle = 0.1 $  \\   
   &  &  (mJy) &   \multicolumn{3}{c}{($10^{34}~{\rm erg~s^{-1}}$)}   \\  \hline 
& \multicolumn{4}{c}{Model 1 (FK06): $\mu = -1.1 $ and $\sigma = 0.9 $ } $N_{10}=688 \pm 82$ & \\ \hline
 47~Tuc  & $ 71 \pm 19 $ & $ 3.1 \pm 0.8 $  &  $  10 \pm 5 $ & $  7.7 \pm 3.6 $ & $  13 \pm 6$  \\ 
 M\,3   &  $ 24 \pm 12 $   &  $ 0.16 \pm 0.08 $ &  $  3.5 \pm 2.2 $ & $  2.6 \pm 1.6 $ & $ 4.3 \pm 2.7  $  \\ 
 M\,5   &  $ 24 \pm 11 $  & $ 0.31 \pm 0.14 $ &  $  3.5 \pm 2.1 $ & $  2.6 \pm 1.6 $ & $ 4.3 \pm 2.6 $  \\ 
 M\,13  &  $ 25 \pm 11 $  & $ 0.38 \pm 0.16 $ &  $   3.6 \pm 2.1 $ & $  2.7  \pm 1.6 $ & $ 4.5 \pm 2.6$  \\ 
 Ter~5  &  $ 167 \pm 33 $  & $ 3.7 \pm 0.7 $ &  $   24 \pm 11 $ & $   18 \pm 8 $ & $ 30  \pm 13$  \\ 
 NGC~6440  & $ 88 \pm 39 $ & $ 0.86 \pm 0.4 $  &  $  13 \pm 8$ & $  10 \pm 6$ & $   16 \pm 9$  \\ 
 NGC~6517  & $ 46 \pm 23 $ & $ 0.29 \pm 0.15 $  &  $   6.6 \pm 4.2$ & $   5.0 \pm 3.1$ & $   8.3 \pm 5.2$  \\ 
 M\,28  & $ 120 \pm 40 $ & $ 2.6 \pm 0.9 $  &  $  17 \pm 9$ & $  13 \pm 7$ & $   22 \pm 11$  \\ 
 NGC~6752  & $ 44 \pm 20 $  & $ 1.7 \pm 0.8 $ &  $  6.3 \pm 3.8 $ & $  4.8 \pm 2.8  $ & $   7.9 \pm 4.7$  \\  
 M\,15  & $ 79 \pm 30 $  & $ 0.52 \pm 0.20 $ &  $  11 \pm 6$ & $   8.5 \pm 4.6 $ & $  14 \pm 8$  \\ \hline 
& \multicolumn{4}{c}{Model 2 (maximum $P_{ks}$): $\mu = -0.61 $ and $\sigma = 0.65 $ } $N_{10}= 453 \pm 56$ & \\ \hline
 47~Tuc  & $ 44 \pm 12 $ & $  2.2 \pm 0.6 $  &  $   6.3 \pm 3.0 $ & $  4.8 \pm 2.2$ & $  7.9 \pm 3.8 $  \\ 
 M\,3   &  $ 15 \pm 8 $   & $  0.11 \pm 0.06 $  &  $   2.2 \pm 1.4 $ & $  1.6 \pm 1.1$ & $  2.7 \pm 1.8$  \\ 
 M\,5   &  $ 15 \pm 7 $  & $ 0.20 \pm 0.09  $ &  $  2.2 \pm 1.3 $ & $   1.6 \pm 1.0 $ & $ 2.7  \pm 1.6 $  \\ 
 M\,13  &  $ 16 \pm 7 $  & $ 0.24 \pm 0.11  $ &  $  2.3 \pm 1.3 $ & $ 1.7 \pm 1.0 $ & $  2.9 \pm 1.7$  \\ 
 Ter~5  &  $ 100 \pm 20 $  & $ 2.6 \pm 0.5  $  &  $ 14 \pm 6 $ & $  11 \pm 5 $ & $  18 \pm 8$  \\ 
 NGC~6440  & $ 68 \pm 30$ & $ 0.75 \pm 0.33  $  & $ 9.8 \pm 5.8$ & $   7.3 \pm 4.3$ & $ 12 \pm 7$  \\ 
 NGC~6517  & $ 30 \pm 15 $ & $ 0.21 \pm 0.10  $  &  $  4.3 \pm 2.7$ & $  3.2 \pm 2.1 $ & $ 5.4 \pm 3.4 $  \\ 
 M\,28  & $85 \pm 28$ & $ 2.0 \pm 0.7  $  &  $  12 \pm 6 $ & $  9.2 \pm 4.7 $ & $  15.3 \pm 7.8 $  \\ 
 NGC~6752  &  $ 27 \pm 12$  &$ 1.1 \pm 0.5  $  &  $ 3.9 \pm 2.3 $ & $  2.9 \pm 1.7$ & $  4.9 \pm 2.9$  \\  
 M\,15  & $53 \pm 20$  & $ 0.40 \pm 0.15  $  &  $ 7.6 \pm 4.1$ & $ 5.7 \pm 3.1$ & $  9.5 \pm 5.2$  \\ \hline 
& \multicolumn{4}{c}{Model 3 (minimum $\chi^2$): $\mu = - 0.52$ and $\sigma = 0.68 $ $N_{10}=354 \pm 43$} & \\ \hline
 47~Tuc  & $ 37 \pm 10$ &  $ 2.5 \pm 0.7  $ &  $  5.3 \pm 2.5 $ & $ 4.0 \pm 1.9$ & $   6.6 \pm 3.1 $  \\ 
 M\,3   & $ 12 \pm 6$    & $ 0.13 \pm 0.06  $  & $ 1.7 \pm 1.1 $ & $  1.3 \pm 0.8 $ & $   2.2 \pm 1.4 $  \\ 
 M\,5   &  $ 13 \pm 6$  & $ 0.24 \pm 0.11  $ &  $  1.9 \pm 1.1$ & $  1.4 \pm 0.8$ & $  2.3 \pm 1.4$  \\ 
 M\,13  &  $ 14 \pm 6$  & $ 0.30 \pm 0.13  $ &  $ 2.0 \pm 1.2$ & $   1.5 \pm 0.9 $ & $  2.5 \pm 1.4$  \\ 
 Ter~5  &  $ 82 \pm 16$  & $ 2.9 \pm 0.6  $ &   $   12 \pm 5 $ & $   8.9 \pm 3.8$ & $  15 \pm 6$  \\ 
 NGC~6440  & $ 48 \pm 21$ & $ 0.74 \pm 0.33  $  &   $  6.9 \pm 4.0$ & $  5.2 \pm 3.0 $ & $ 8.6 \pm 5.1$  \\ 
 NGC~6517  & $ 23 \pm 12 $ & $ 0.21 \pm 0.11  $  &  $   3.3 \pm 2.1$ & $ 2.5 \pm 1.6$ & $  4.1 \pm 2.7$  \\ 
 M\,28  & $ 63 \pm 21 $ & $  1.5 \pm 0.5 $ &  $  9.1 \pm 4.6$ & $  6.8 \pm 3.5$ & $  11 \pm 6$  \\ 
 NGC~6752  & $ 21 \pm 10$  & $  1.2 \pm 0.5 $ &  $  3.0  \pm 1.8 $ & $  2.3 \pm 1.4 $ & $ 3.8 \pm 2.3$  \\  
 M\,15  &  $ 41 \pm 15$ &  $ 0.43 \pm 0.16  $ &  $   5.9 \pm 3.2 $ & $  4.4 \pm 2.4$ & $  7.4  \pm 3.9$ \\ \hline \hline
\end{tabular}
\label{tb:predlnorm}
\end{table*}

\begin{table*}\footnotesize
\caption{Population estimations and predictions using power-law luminosity
functions. See Table 2 for details about tabulated parameters.}
\begin{tabular}{|l|c|c|c|c|c|} \hline \hline 
Cluster  & $N_{\rm rad}$  & $S_{\rm sim, tot}$  &  \multicolumn{3}{c}{$L_{\rm \gamma,sim}$}   \\  
  &  &   & $ \langle \eta_{\gamma} \rangle = 0.08 $  & $ \langle \eta_{\gamma} \rangle = 0.06 $ &  $ \langle \eta_{\gamma} \rangle = 0.1 $  \\   
   &  &  (mJy) &   \multicolumn{3}{c}{($10^{34}~{\rm erg~s^{-1}}$)}   \\  \hline 
  & \multicolumn{4}{c}{Model 4: $L_{\rm min} = 0.017 $ and $\beta = 0.92 $ $N_{10} = 3399 \pm 421$ } & \\ \hline
 47~Tuc  & $ 313 \pm 84 $ &  $ 3.4 \pm 0.89  $  &  $ 45 \pm 21   $ & $ 34 \pm 16  $ & $ 56 \pm 27  $  \\  
 M\,3   &  $ 114 \pm 57 $  &  $ 0.19 \pm 0.09  $   & $ 16 \pm 10  $ & $ 12 \pm 8  $ & $ 21 \pm 13  $  \\  
 M\,5   &  $ 112 \pm 50 $  & $ 0.32 \pm 0.14 $   & $ 16 \pm 10  $ & $  12 \pm 7  $ & $ 21 \pm  12  $  \\  
 M\,13  &  $ 118 \pm 52 $  & $ 0.40 \pm 0.18  $  & $ 17 \pm 10  $ & $ 13 \pm 8  $ & $ 21 \pm 13  $  \\  
 Ter~5  & $ 764 \pm 153 $  & $ 4.4 \pm  0.9  $  & $ 110 \pm 48  $ & $ 83 \pm 36  $ & $ 138 \pm 60  $  \\  
 NGC~6440  & $485 \pm 217  $ &  $ 1.2 \pm 0.5  $  & $ 70 \pm 41 $ & $ 53 \pm 31  $ & $ 87 \pm 52  $  \\  
 NGC~6517  & $ 238 \pm 119   $ & $ 0.35 \pm 0.18  $  &  $ 34 \pm 22  $ & $ 26 \pm 16  $ & $ 43 \pm 27  $  \\  
 M\,28  & $ 628 \pm 209 $ &   $ 2.5 \pm 0.8  $   &  $ 90 \pm 46  $ & $ 68 \pm 35 $ & $ 113 \pm 58  $  \\  
 NGC~6752  & $216 \pm 97  $  &  $ 1.9 \pm 0.9  $   & $ 31 \pm 19  $ & $ 23 \pm 14  $ & $  39 \pm 23  $  \\   
 M\,15  & $ 411 \pm 155  $  &  $  0.68 \pm 0.25  $   & $  59 \pm 32   $ & $ 449 \pm 24  $ & $74 \pm 40  $  \\ \hline
  & \multicolumn{4}{c}{Model 5: $L_{\rm min} = 0.022 $ and $\beta = 1.01  $ $N_{10} = 3767 \pm 478$ } & \\ \hline
 47~Tuc  & $ 324 \pm 87  $ & $ 3.3 \pm 0.9  $   &  $ 47 \pm 22    $ & $  35 \pm 17   $ & $  58 \pm 28   $  \\  
 M\,3   & $ 121 \pm 61  $   &  $ 0.19 \pm 0.09  $   & $ 17 \pm 11  $ & $ 13 \pm 8  $ & $ 22 \pm 14  $  \\  
 M\,5   &  $ 116 \pm 52 $  & $ 0.31 \pm 0.14  $   & $ 17 \pm 10  $ & $ 13 \pm 7  $ & $ 21 \pm 12  $  \\  
 M\,13  &  $ 123 \pm 55 $  & $ 0.39 \pm 0.18  $   & $ 18 \pm 11  $ & $  13 \pm 8  $ & $ 22 \pm 13  $  \\  
 Ter~5  &  $815 \pm 163 $ & $ 4.5 \pm 0.9  $  & $ 117 \pm 51  $ & $ 88 \pm 39  $ & $ 147 \pm 64  $  \\  
 NGC~6440  & $ 580 \pm 260 $ &  $ 1.3 \pm 0.6  $  & $ 84 \pm 50  $ & $ 63 \pm 37  $ & $  104 \pm 62  $  \\  
 NGC~6517  & $ 271 \pm 136 $ & $ 0.38 \pm 0.20  $  &  $ 39 \pm 25 $ & $ 29 \pm 19  $ & $ 49 \pm 31  $  \\  
 M\,28  & $ 714 \pm 238 $ &  $ 2.7 \pm 0.9  $   &  $ 103 \pm 53  $ & $ 77 \pm 30  $ & $ 129 \pm 66  $  \\  
 NGC~6752 &  $ 237 \pm 106 $ &  $ 2.0 \pm 0.9 $   & $ 34 \pm 20  $ & $ 26 \pm 15  $ & $ 43 \pm 25  $  \\   
 M\,15  & $ 466 \pm 176  $  &  $ 0.73 \pm  0.27  $  & $ 67 \pm 36  $ & $ 50 \pm 27  $ & $ 84 \pm 46  $  \\ \hline 
   & \multicolumn{4}{c}{Model 6 (FG00): $L_{\rm min} = 0.03 $ and $\beta = 0.85 $ $N_{10} = 1590 \pm 194$} & \\ \hline
 47~Tuc  & $ 153 \pm 41   $ & $  3.2 \pm 0.86  $   &  $ 22 \pm  10  $ & $  17  \pm 8    $ & $ 28 \pm 13   $  \\  
 M\,3   &  $ 55 \pm 28   $  & $ 0.18 \pm 0.09 $    & $ 7.9 \pm  5.1  $ & $ 5.9  \pm 3.8    $ & $ 9.9 \pm 6.3 $  \\  
 M\,5   &  $ 54 \pm 24   $  & $ 0.31 \pm 0.14 $  & $ 7.8 \pm  4.6  $ & $ 5.8  \pm 3.4   $ & $ 9.7 \pm 5.7  $  \\  
 M\,13  &   $ 57 \pm 26   $ & $ 0.38 \pm 0.17 $   & $  8.2 \pm  4.9  $ & $ 6.2  \pm 3.7  $ & $ 10 \pm 6   $  \\  
 Ter~5  & $ 363 \pm 73   $  & $ 4.2 \pm 0.8 $  & $ 52 \pm  23 $ & $ 39  \pm 17   $ & $ 65 \pm 29    $  \\  
 NGC~6440  & $ 216 \pm 96   $ &  $ 1.0 \pm 0.5 $  & $ 31 \pm  18 $ & $ 23  \pm 14   $ & $ 39 \pm 23   $  \\  
 NGC~6517  & $  112 \pm 56     $ &  $ 0.33 \pm 0.16 $ &  $ 16 \pm  10  $ & $ 12  \pm 8    $ & $  21 \pm 13  $  \\  
 M\,28  &  $ 287 \pm 95    $ & $ 2.2 \pm 0.7 $  &  $ 41 \pm  21  $ & $ 31  \pm 16    $ & $  52 \pm 26   $  \\  
 NGC~6752  & $ 101 \pm 45 $  &  $ 1.8 \pm 0.8 $   & $ 15 \pm  9  $ & $ 11  \pm 6  $ & $ 18 \pm 11    $  \\   
 M\,15  & $ 192 \pm 73   $  & $ 0.62 \pm 0.23 $ & $ 28 \pm  15 $ & $  21  \pm 11  $ & $ 35 \pm 19  $  \\ \hline \hline
\end{tabular}
\label{tb:predplawLmax}
\end{table*}

%%%%%%%%%%%%%%%%%%%%%%%%%%%%

\begin{table*}\footnotesize
\caption{Observed and inferred properties of globular clusters containing pulsars used in the present work. From left to right, we list the GC name, distance from the Sun $d$ and from the Galactic center $d_{\rm gcen}$ (both in kpc), concentration parameter $c$ (base 10 logarithm of the ratio of the tidal radius to core radius), core radius $r_c$ (pc), base-10 logarithm of the central density ($\rho_c$ in solar luminosities per cubic parsec),  velocity dispersion $v_c$ (km~s$^{-1}$), base-10 logarithms of the cluster mass ($M_{GC}$ in solar masses) and base-10 logarithms of the core relaxation timescale ($t(r_c)$ in yr), metallicity [Fe/H], normalized two body encounter rate $\Gamma_{\rm norm}$, gamma-ray flux $L_{\gamma}$ ($10^{34}$~erg~s$^{-1}$) and inferred number of gamma-ray pulsars $N_{\gamma}$ from \citet{abd10} and \citet{tkhcll11}. For $\Gamma_{\rm norm}$, we first calculate the two body encounter rate for each GC as $\Gamma=\rho_c^{1.5} \, r_c^2$. The values of core radius have been calculated as $r_c = d \, {\rm tan} \, \theta_c$ where $\theta_c$ are the angular radii as quoted in the latest version of the Harris catalog \citep[][updated in December 2010]{har96}. The central cluster density $\rho_c$, has been calculated as $\rho_c=\Sigma_{c}/(r_c \, p)$ using the values of central surface brightness $\mu_{Vc}$ (in V magnitude per square arcsecond) and the extinction coefficient $A_{V} = 3.1 \, E(B-V)$ where $E(B-V)$ is the color excess. $\Sigma_{c}$, the central surface brightness in $L_{V\odot} \, {\rm pc^2}$ can be calculated as $\log(\Sigma_{c})=0.4 \, [26.392-(\mu_{Vc}-A_{V})]$ and $p$ is a parameter defined as $\log(p)=-0.603 \times 10^{-c}+0.302$ \citep{djor93}. Finally we normalized $\Gamma$ to $\Gamma_{\rm norm}$ considering $\Gamma = 100$ for M62 \citep[following][]{abd10}. The tabulated values for $v_c$ and $\log(M_{\rm GC})$ can be found at \textit{http://www.astro.lsa.umich.edu/$\sim$ognedin/gc/vesc.dat} \citep{gzp02}. 
}
\begin{tabular}{|l|c|c|c|c|c|c|c|c|c|c|c|c|} \hline \hline 
GC & $d$ & $d_{\rm gcen}$ & $c$ & $r_c$ & $\log(\rho_c)$ & $v_c$ & $\log(M_{\rm GC})$ & 
$\log(t(r_c))$ & [Fe/H]& $\Gamma_{\rm norm}$& $L_{\gamma}$ &$N_{\gamma}$ \\ 
\hline 
47~Tuc & 4.03 & 7.4 & 2.07 & 0.42 & 4.93 & 16.4 & 6.17 & 7.84 & --0.72 & 41.27 & $4.8^{+1.1}_{-1.1}$ & $33^{+15}_{-15}$ \\
M\,3 & 10.23 & 12.0 & 1.89 & 1.10 & 3.58 &  9.2 & 5.98 & 8.31 & --1.50 & 2.61 & --- & --- \\
M\,5 & 7.76 & 6.2 & 1.73 & 0.99 & 3.87 &  11.8 & 5.93 & 8.28 & --1.29 &  5.83
& --- & --- \\
M\,13 & 7.13 & 8.4 & 1.53 & 1.28 & 3.55  & 10.3 & 5.89 & 8.51 & --1.53 & 3.26 & --- & --- \\
Ter~5 & 5.50 & 1.2 & 1.62 & 0.25 & 5.26 &  12.7 & 5.57 & 7.57 & --0.23 & 46.50 & $25.7^{+9.4}_{-8.8}$ & $180^{+100}_{-90}$ \\
NGC~6440 & 8.47 & 1.3 & 1.62 & 0.34 & 5.23  & 21.6 & 5.91 & 7.60 & --0.36 & 78.27
  & $19.0^{+13.1}_{-5.0}$ & $130^{+100}_{-60}$ \\
NGC~6517 & 10.60 & 4.2 & 1.82 & 0.19 &  5.30 & 20.6 & 5.72 & 6.92 & --1.23
 & 29.71 & --- & ---\\
 M\,28 & 5.70 & 2.7 & 1.67 & 0.40 & 4.86  & 16.3 & 5.74 & 7.62 & --1.32 & 28.19 & $6.2^{+2.6}_{-1.8}$ & $43^{+24}_{-21}$ \\
 NGC~6752 &  4.42 & 5.2 & 2.50 & 0.22 & 5.01  & 7.1 & 5.50 & 6.88 & --1.54 
& 14.72 & $1.4^{+0.7}_{-0.7}$  & $10^{+15}_{-6}$ \\
M\,15 & 10.30 & 10.4 & 2.29 & 0.42 & 5.08  & 13.4 & 6.08 & 7.84 & --2.37 & 66.81
 & $< 5.8$ & $<56$ \\ \hline \hline 
\end{tabular}
\label{tb:gammaonly}
\end{table*}

\subsection{Diffuse radio emission}
\label{sec:diffuseradio}

A potentially very useful additional constraint comes from
observations of the diffuse radio emission in GCs. Assuming that the
only contribution to this flux is from the pulsars, then such
measurements constrain the integrated luminosity function in a given
cluster. Our Monte Carlo models make specific predictions for these
observations (see Eqn. \ref{eq:flux_pred}).  For Terzan 5, the total
radio flux $S_{\rm obs, tot} = 5.2 ~{\rm mJy~kpc^2}$ (sum of diffuse
flux and the fluxes of point sources) found by \citet{fg00}. 
\citet{fg00} observed some other clusters too, among which NGC~6440 belongs to our list (Table
\ref{tb:gammaonly}). But as they mentioned that their
observation in this cluster is consistent with the position of a
single pulsar PSR B1745$-$20, we can not use this datum for our
study. For 47~Tuc, \citet{mdca04} found\footnote{This sum is essentially equivalent
to the individual fluxes of the 14 pulsars in this cluster with 
measured fluxes so far (see Table 1). The remaining 9 currently
known pulsars, must therefore contribute much less than a mJy
of diffuse flux. For example a typical flux of 30 $\mu$Jy per 
pulsar would bring the diffuse flux to $\sim 2.3$~mJy.} that $S_{\rm obs, tot} = 
2.0 \pm 0.3 ~{\rm mJy~kpc^2}$.

Assuming both the diffuse flux measurements for Terzan~5 
and 47~Tuc are dominated by their respective pulsar populations, we can 
confront them with the
predictions from our simulations. An inspection of Tables \ref{tb:predlnorm} and \ref{tb:predplawLmax} shows
that the observed diffuse flux for 47~Tuc is successfully reproduced
by all of the models, within the nominal uncertainties. For Terzan~5, the
power law models provide a better match to the diffuse flux overall,
while the log-normal models predict a slightly smaller flux that lies
2--5$\sigma$ below the nominal value found by \citet{fg00}.

\begin{figure}
\centerline{\psfig{figure=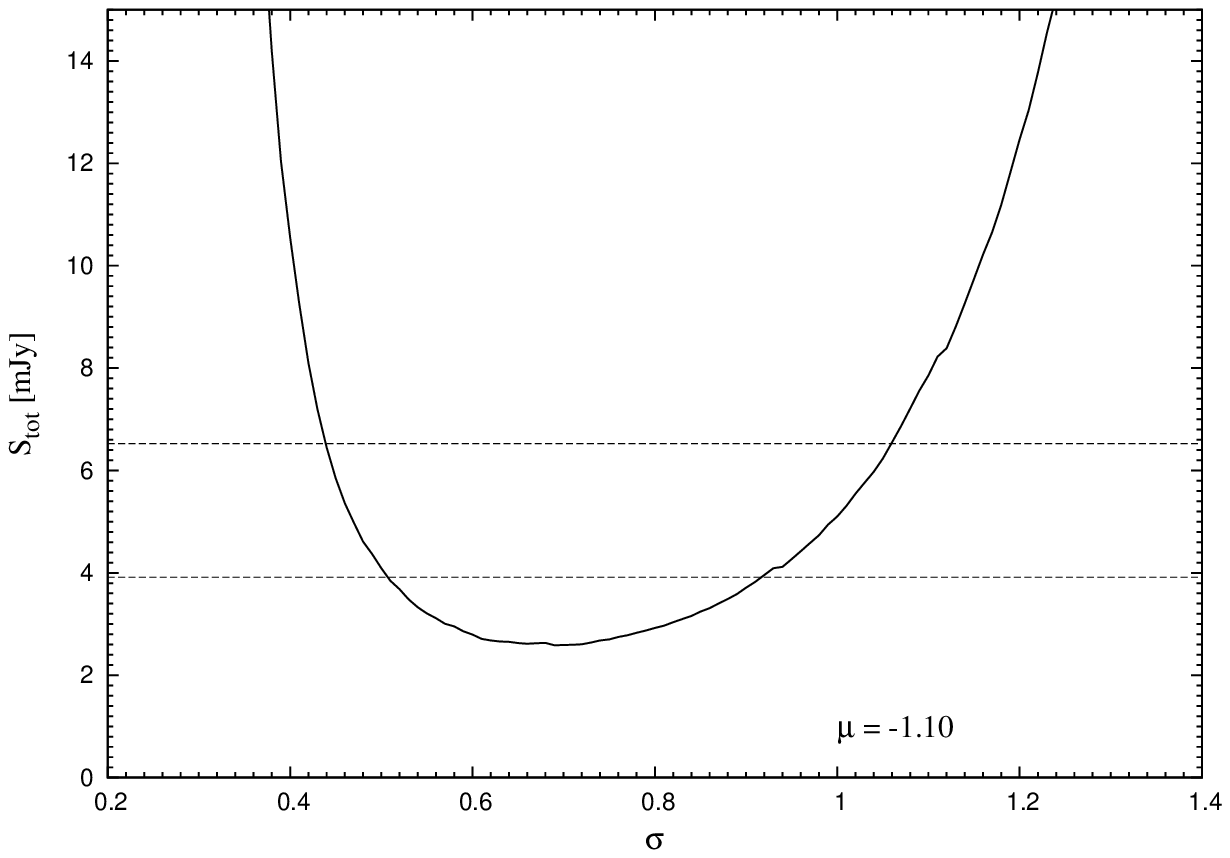,width=7cm,angle=0}}
\centerline{\psfig{figure=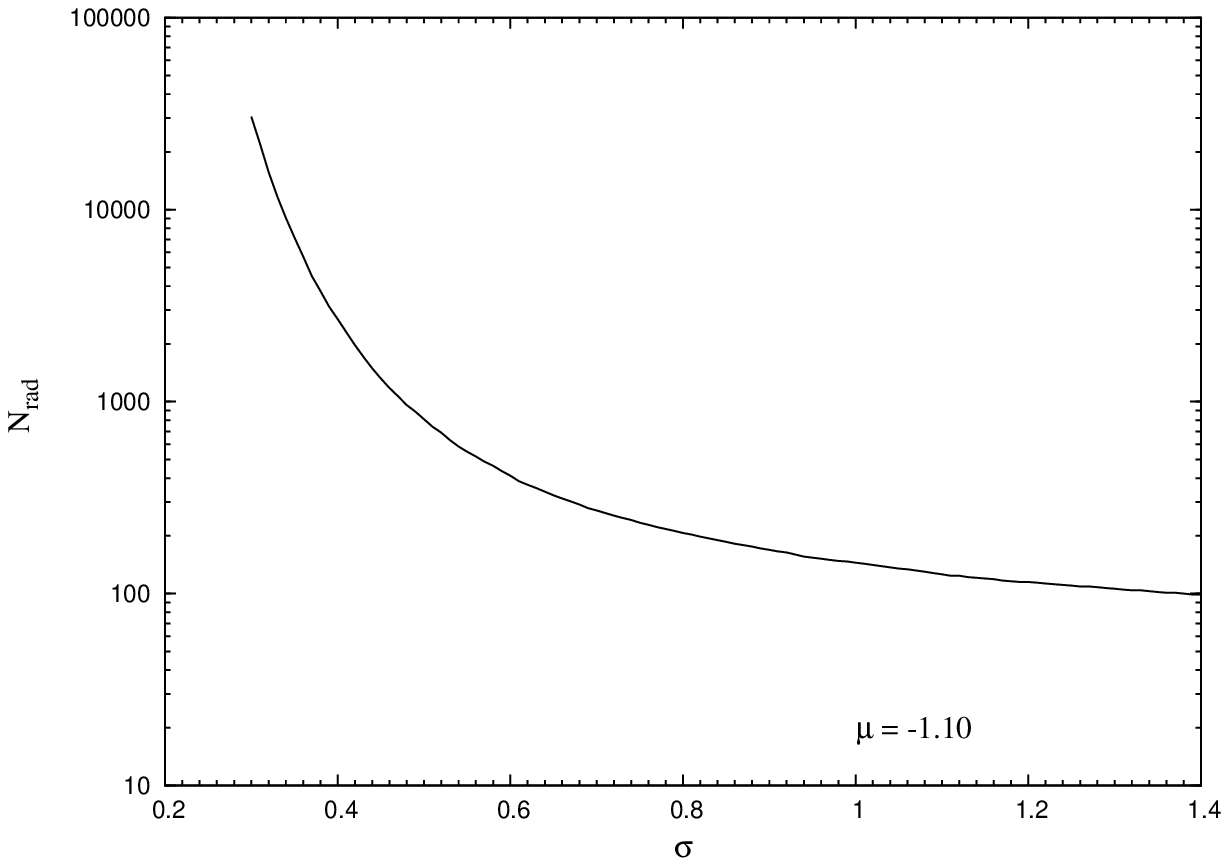,width=7cm,angle=0}}
\caption{Variations of total radio flux (upper panel) and predicted number of pulsars (lower panel) in Terzan 5 as obtained from our simulations with log-normal distribution keeping $\mu$ fixed at $-1.1$ and varying $\sigma$. The dashed lines in the upper panel denote the ranges of $S_{\rm sim, tot} = S_{\rm obs, tot} \pm 25\%$.} \label{fig:SNtotter5_lognormsigvar}
\end{figure} 

In Fig.~\ref{fig:SNtotter5_lognormsigvar}, we fix $\mu$ to the nominal
value from model 1, i.e.~as found by FK06 for normal pulsars (--1.1) and vary $\sigma$. 
For this case, we see that there are only two possible
ranges of $\sigma$ which are compatible with the diffuse flux
measurement of Terzan 5: $\sigma \sim 0.5$ or $\sigma \sim 0.9$ (see the upper panel of Fig.~\ref{fig:SNtotter5_lognormsigvar}). The ``solution''
with $\sigma \sim 0.5$, however lies well outside the $\chi^2$
contours shown in Fig.~\ref{fig:kschisq_lognorm}. We therefore favor
the region with $\mu \sim -1.1$ and $\sigma \sim 0.9$ (the nominal
FK06 values) which is consistent with both constraints.  In this case,
the implied total number of pulsars $N_{\rm tot} \sim 150$ (see the lower panel of Fig.~\ref{fig:SNtotter5_lognormsigvar}).  Further
constraints on these parameters using a more detailed Bayesian
analysis of these constraints for Terzan~5 will be the subject of a
subsequent paper (Chennamangalam et al.~in preparation).

\begin{figure}
\centerline{\psfig{figure=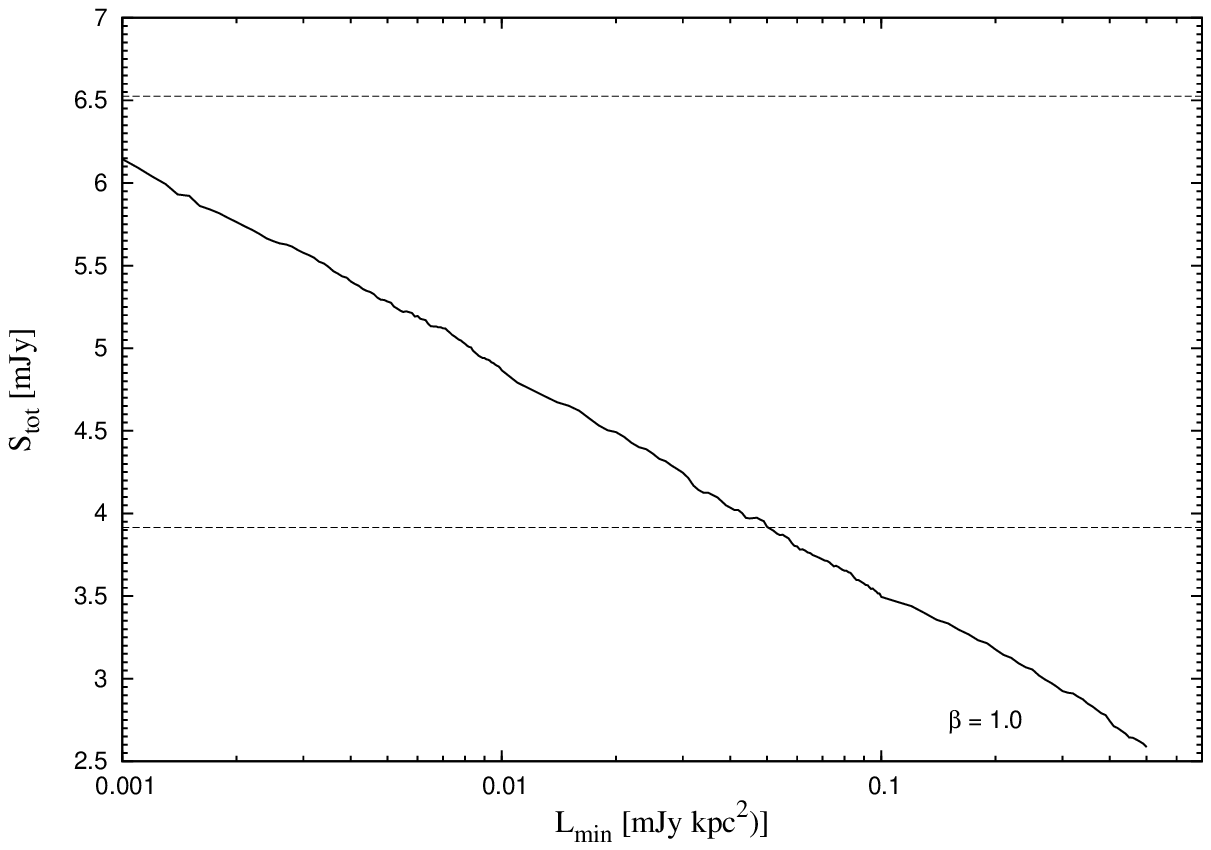,width=7cm,angle=0}}
\centerline{\psfig{figure=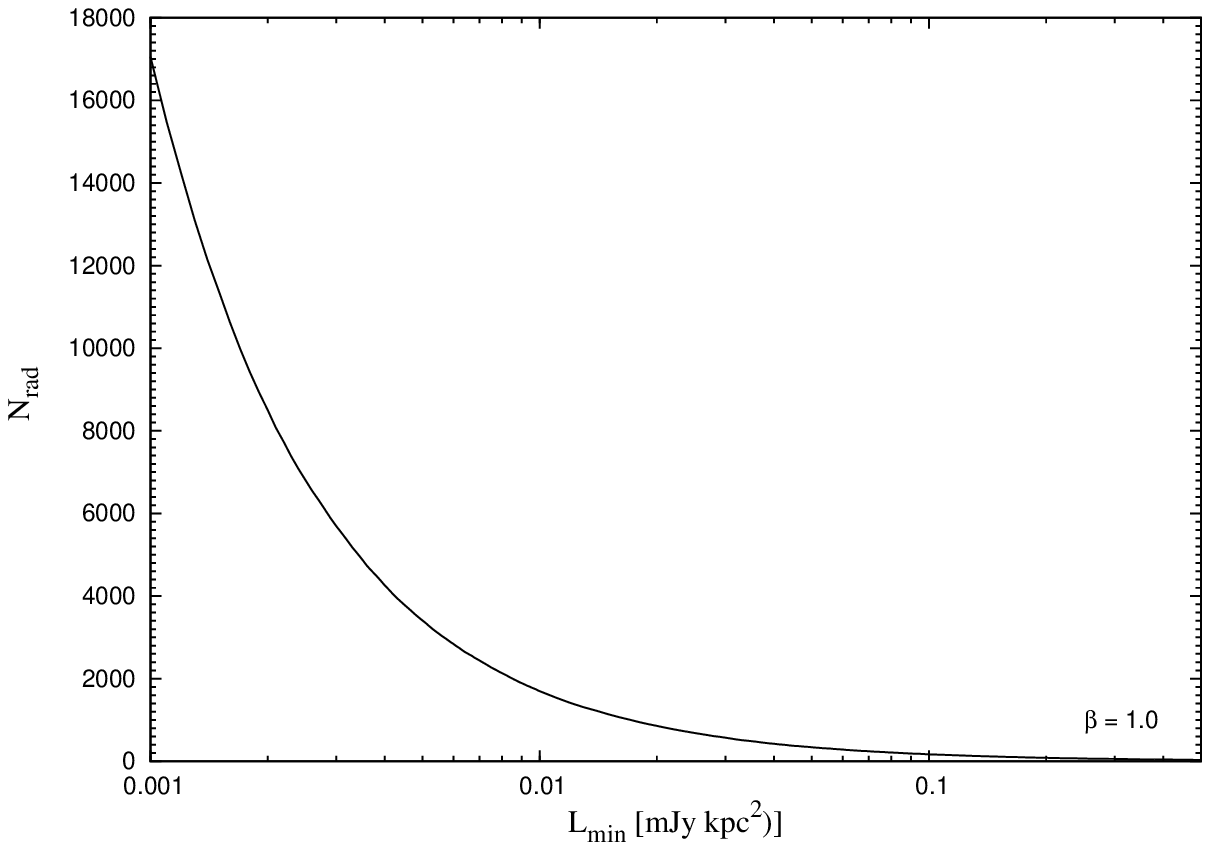,width=7cm,angle=0}}
\caption{Variations of total radio flux (upper panel) and predicted number of pulsars (lower panel) in Terzan 5 as obtained from our simulations with Power law distribution keeping $\beta$ fixed at $1.0$ and varying $L_{\rm min}$. The dashed lines in the upper panel denote the ranges of $S_{\rm sim, tot} = S_{\rm obs, tot} \pm 25\%$.} \label{fig:SNtotter5_powlawbetvar}
\end{figure}

Fig.~\ref{fig:SNtotter5_powlawbetvar} shows the analogous diagram to
Fig.~\ref{fig:SNtotter5_lognormsigvar} for the power-law luminosity
function for the choice $\beta=1$. In this case, there is a wide range
of $L_{\rm min}$ values that are consistent with the diffuse flux
measurements (upper panel of Fig.~\ref{fig:SNtotter5_powlawbetvar}), and no significant additional constraints on $L_{\rm
  min}$ can be made. The upper bound of $S_{\rm tot}$ gives an extremely high value of $N_{\rm tot} \sim 1000$ which seems unrealistic, but the lower bound of $S_{\rm tot}$ gives $N_{\rm tot}=340$ for $L_{\rm min} = 0.05~{\rm mJy~kpc^2}$ (lower panel of Fig.~\ref{fig:SNtotter5_powlawbetvar}). 

\subsection{Predicted population sizes and diffuse fluxes for different GCs} 
\label{sec:diffusegamma}

The detection of diffuse gamma-ray emission from GCs has allowed
some constraints to be placed on $N_{\gamma}$, the number of
gamma-ray emitting pulsars in each cluster. Following \citet{abd10},
we can write the total gamma-ray luminosity
\begin{equation}
L_{\gamma} = N_{\gamma} \langle \dot{E} \rangle \, \langle \eta_{\gamma} \rangle,
\label{eq:ngamma}
\end{equation} 
where $\langle \dot{E} \rangle$ is the average spin-down power of
MSPs, $\langle \eta_{\gamma} \rangle$ is the average spin-down to
gamma-ray luminosity conversion efficiency. As the values of $\langle
\dot{E} \rangle$ and $\langle n_{\gamma} \rangle$ are not well known,
\citet{abd10} assumed $\langle \dot{E} \rangle = (1.8 \pm 0.7) \times
10^{34}~{\rm erg~s^{-1}}$ and $ \langle \eta_{\gamma} \rangle =
0.08$. We use the values of $N_{\gamma}$ estimated using the above
relationship for the clusters with gamma-ray flux (and hence
luminosity) measurements.

\begin{figure}
\centerline{\psfig{figure=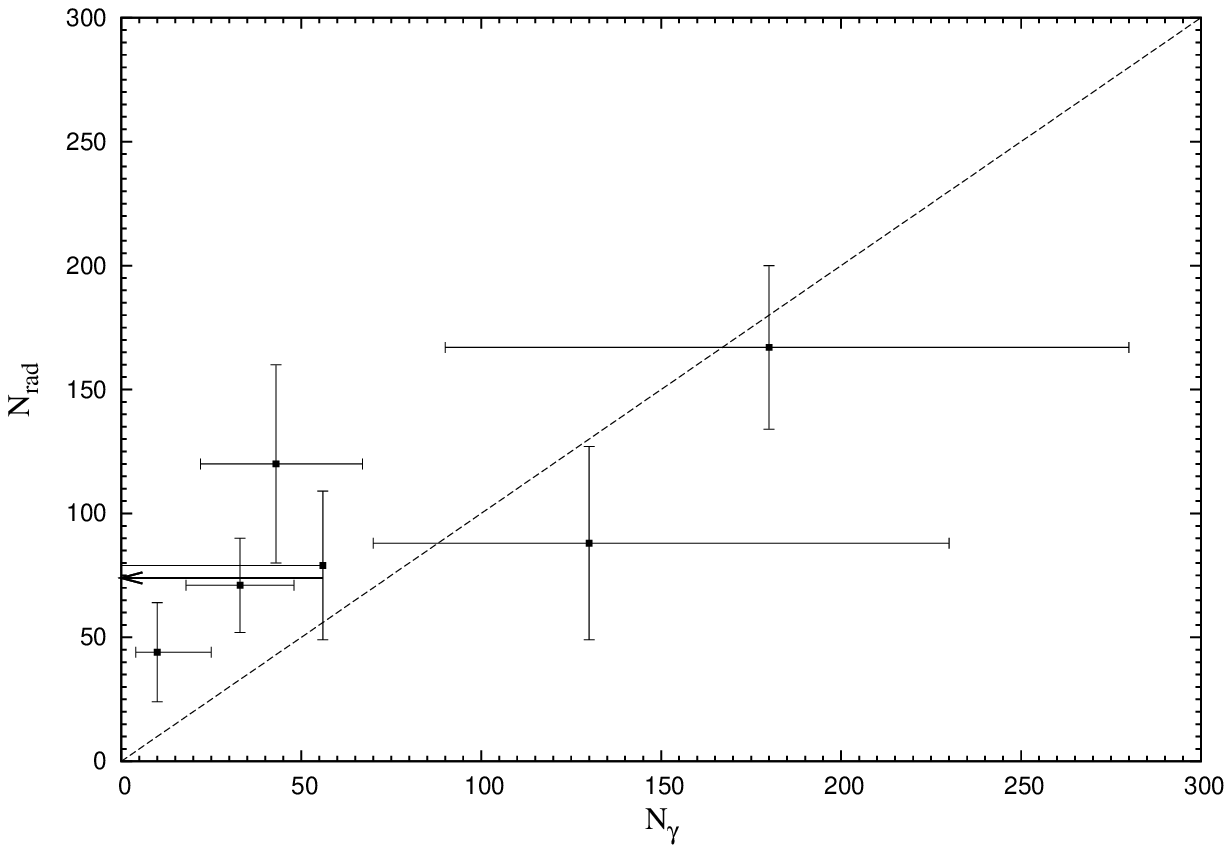,width=7cm,angle=0}}
\centerline{\psfig{figure=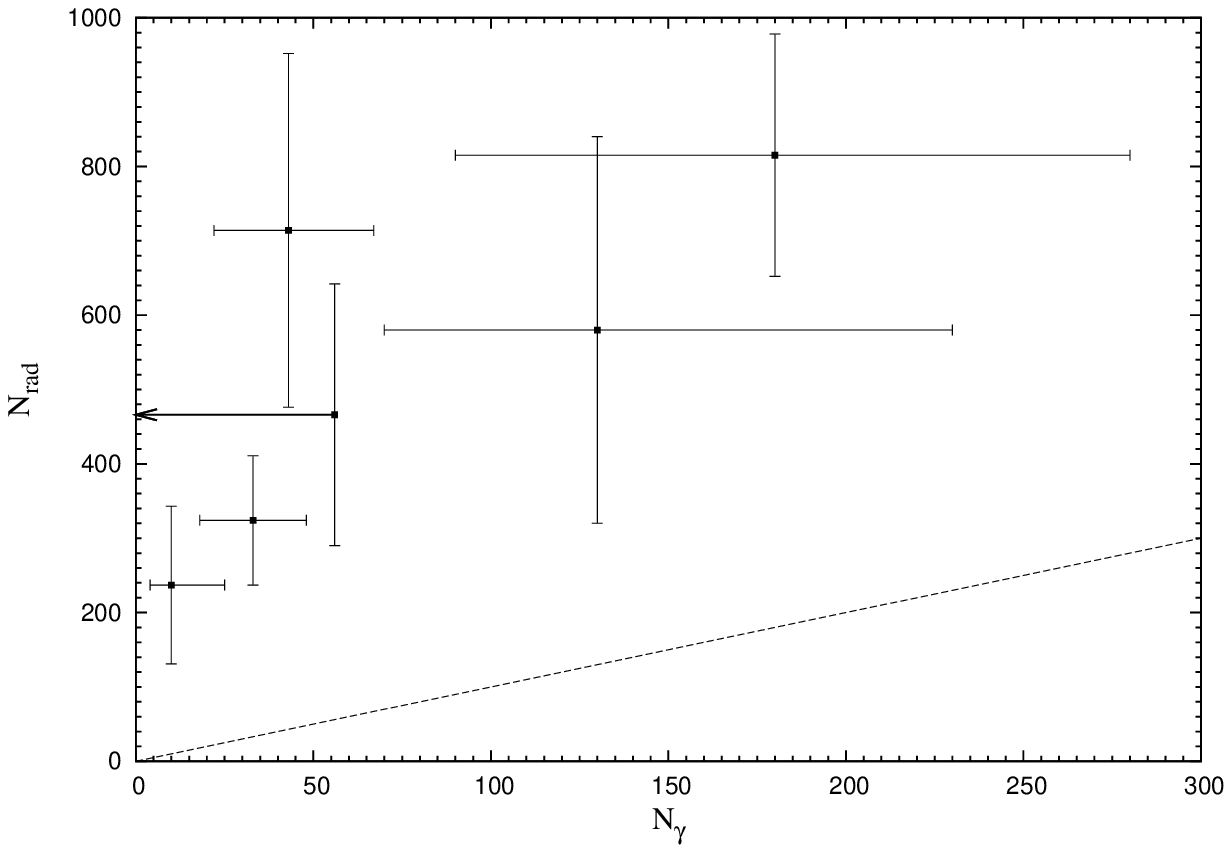,width=7cm,angle=0}}
\caption{Comparison of predicted number of pulsars from our simulations with those obtained from $\gamma$-ray fluxes. Top: log-normal luminosity function (FK06 parameters). Bottom: power-law luminosity function (Model 5).} \label{fig:npsrcompare}
\end{figure}  

In Fig.~\ref{fig:npsrcompare}, we
compare the estimates of $N_{\gamma}$ with our predicted
numbers of radio pulsars ($N_{\rm rad}$) for one model of each luminosity function.
A reasonable agreement can be noted for the log-normal function, but for the power-law function, the values of $N_{\rm rad}$ are significantly larger than those of $N_{\gamma}$. This fact remains unchanged even if we choose other models from these luminosity functions (see Tables \ref{tb:predlnorm} and \ref{tb:predplawLmax}). Although this simple analysis does provide some support to the log-normal models, due to the assumptions made in equation (\ref{eq:ngamma})
and implicitly assuming that $N_{\gamma}=N_{\rm rad}$,
it does not help to constrain their values significantly.

By assuming $N_{\gamma} = N_{\rm rad}$, we obtain $L_{\gamma, {\rm sim}}$ from
equation (\ref{eq:ngamma}). We tabulate $L_{\gamma, {\rm sim}}$ for different
choices of $\langle \eta_{\gamma} \rangle$ for different models in
Tables \ref{tb:predlnorm} and \ref{tb:predplawLmax}. These values can
be compared with observed values of $L_{\gamma}$ and $N_{\gamma}$ as
shown in Table \ref{tb:gammaonly}. 
It is apparent that the $\gamma$-ray luminosities
predicted by the power-law models are generally higher than observed.
We note, however, that in addition to the explicit assumptions
about beaming geometry mentioned above, it has been recently shown
that the $\gamma$-ray observations can be biased by one or more
very bright pulsars in the cluster \citep{paulo11}, and may not be
representative of the diffuse flux of the whole population.

\section{Comparison with earlier results} 
\label{sec:comp}

\begin{figure}
\centerline{\psfig{figure=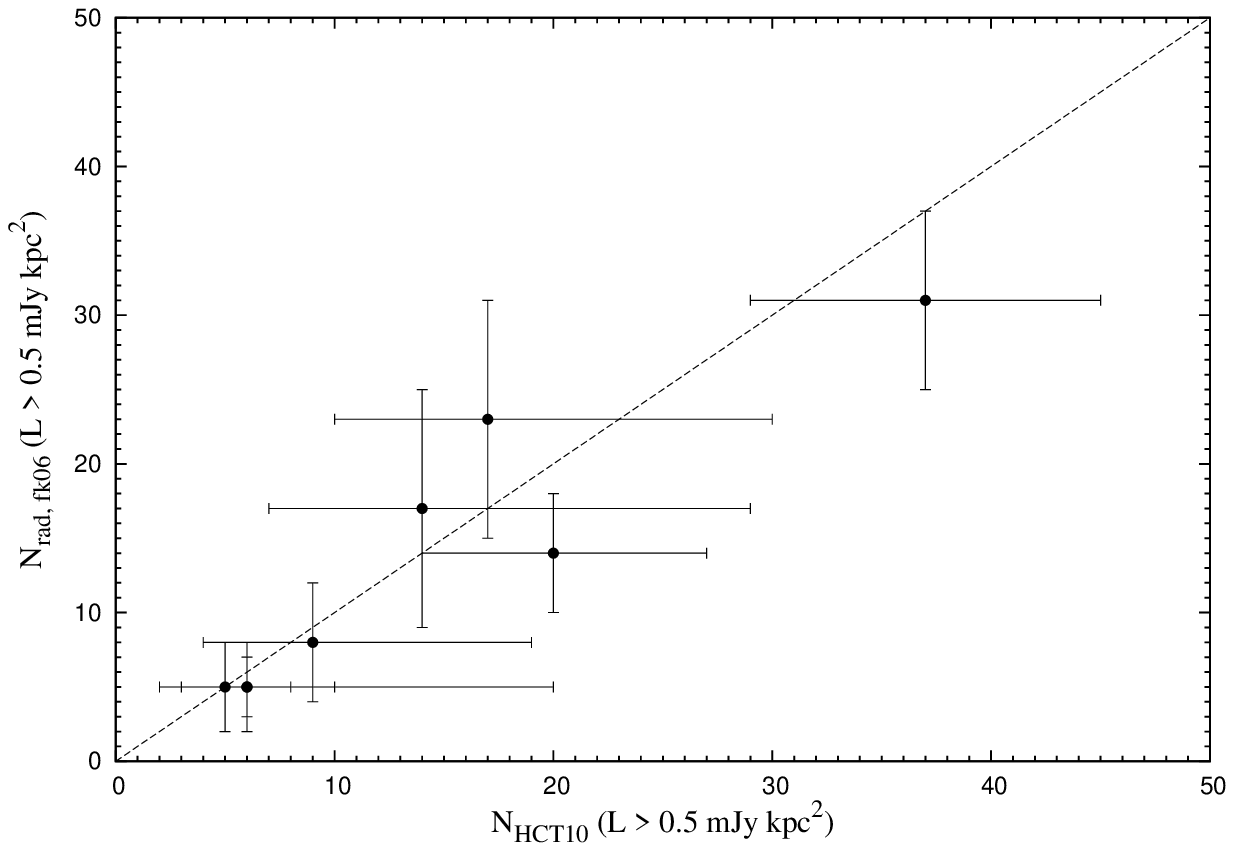,width=7cm,angle=0}}
\caption{Comparison of our predicted number of pulsars in different GCs using FK06 parameters to those by HCT10.} \label{fig:comp_hctkf06}
\end{figure}

In Fig.~\ref{fig:comp_hctkf06}, we compare our predicted number of
pulsars having $L > 0.5 ~{\rm mJy~kpc^2}$ in different GCs using FK06
parameters to those by HCT10. For Terzan~5, we do not use the distance
they adopted for this cluster (10.3 kpc). Instead, here we re-calculate the value of $N(L >
0.5)$ by adopting exactly the same method as HCT10 using the recent
estimate 5.5 kpc \citep{ort07} to calculate luminosities. The overall
agreement is good which again highlights the fact that,
at least above 0.5~mJy~kpc$^2$, the exact form
of the luminosity function for GC pulsars is not uniquely specified by
the current sample of luminosities.

\begin{figure*}
\centerline{\psfig{figure=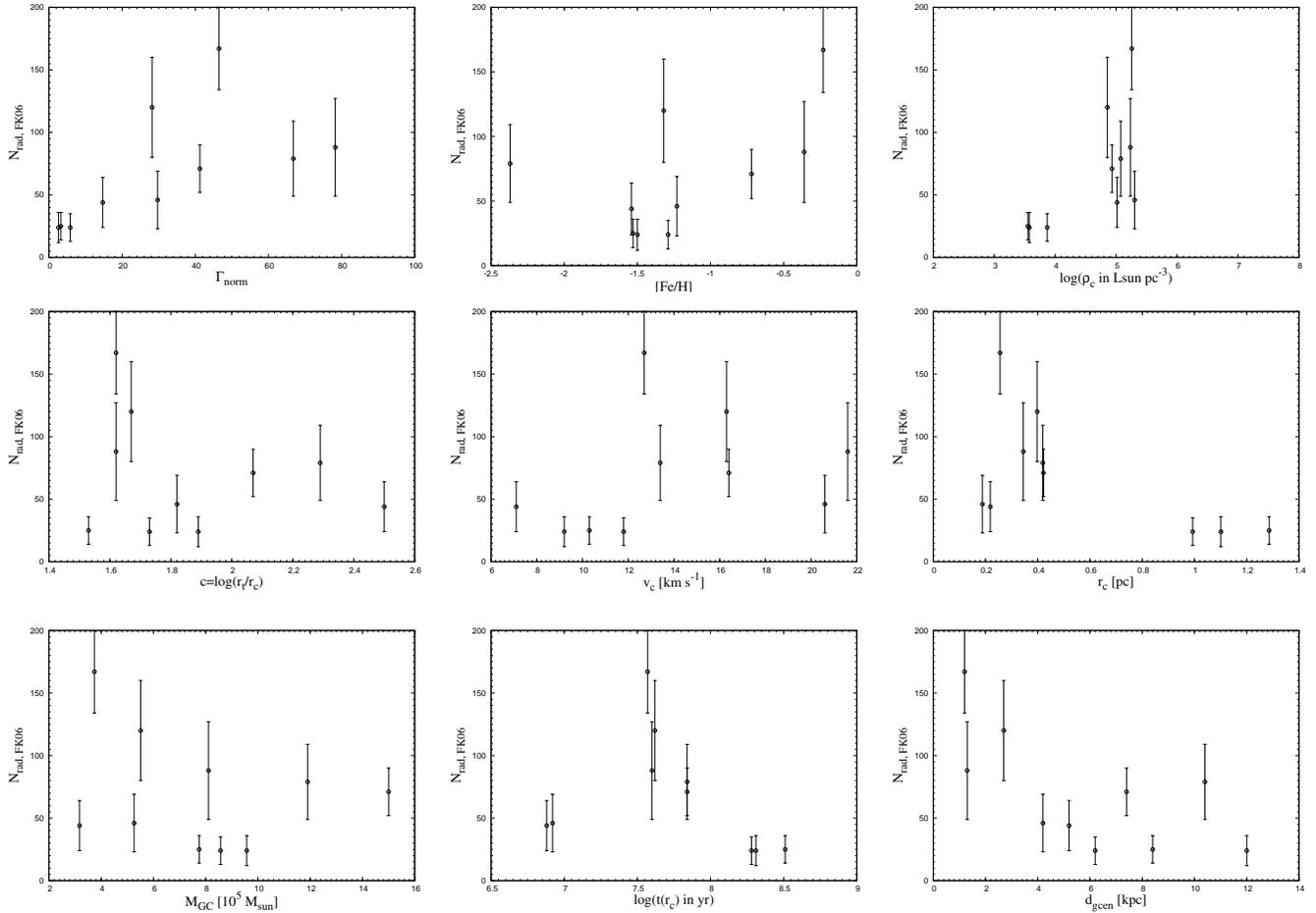,width=\textwidth,angle=0}}
\caption{Plot of predicted number of pulsars in different GCs using
  FK06 parameters against various cluster parameters (see text).
  } \label{fig:corrln_withGcparams}
\end{figure*}

HCT10 used their power-law luminosity functions to search
for correlations between the number of inferred radio pulsars and
fundamental cluster parameters. In their Fig.~3, they present evidence
for a correlation between $N_{\rm rad}$ and both cluster
metallicity $[$Fe/H$]$ as well as the two-body encounter rate
$\Gamma_{\rm norm}$. An inspection of these diagrams suggests that
the claimed correlations are strongly influenced by Terzan~5.
Adopting, for the purposes of this discussion, the parameters of model 1
(i.e.~the log-normal luminosity parameters found by FK06), together
with the revised distance to Terzan 5, we revisit these
proposed correlations in Fig.~\ref{fig:corrln_withGcparams}.  
Also shown here are the results of
correlation tests between $N_{\rm rad}$ and other cluster parameters. We searched for relationships between the distance
of each GC and the galactic center, $d_{\rm gcen}$, the logarithm of the central
luminosity density, $\rho_c$, the concentration parameter\footnote{This
parameter is defined to be the logarithm of the ratio of the GC's
tidal radius to its core radius.}, $c$, the logarithm of the core relaxation time
$t(r_c)$, the cluster mass, $M_{\rm GC}$, the central
velocity dispersion, $v_c$, and the core radius, $r_c$. The
parameter values used for this analysis are given in Table~\ref{tb:gammaonly}.

As can be seen in Fig.~\ref{fig:corrln_withGcparams}, none of the
scatter diagrams provides compelling evidence for a direct relationship
between $N_{\rm rad}$ and any cluster parameters.
The lack of any statistically significant correlations can also be seen
formally in Table~\ref{tb:stat}, where we have calculated Pearson's correlation
coefficient ($rp$) and probability at which the null hypothesis of
zero correlation is disproved ($P_{rp}$), Spearman correlation
coefficient ($rs$) and probability at which the null hypothesis of
zero correlation is disproved ($P_{rs}$), Kendall's $\tau$ and the
probability at which the null hypothesis of zero correlation is
disproved ($P_{\tau}$). 

We have shown the results of correlation analyses only for FK06 (our model
1), but for other models the results are almost the same.  Because
both Spearman correlation and Kendall's $\tau$ test are based on
ranks, and in all the models, the GCs with descending order of ranks
(based on $N_{\rm rad}$) are as follows Terzan 5, M\,28, NGC 6440,
M\,15, 47~Tuc, NGC~ 6517, NGC~6752, M\,13, M\,3 and M\,5 (with the
exception in model 1 and 2 where there is a tie between M\,3 and M\,5,
but the order of other GCs are the same, see tables \ref{tb:predlnorm} and \ref{tb:predplawLmax}). Even for the
parametric test - Pearson's correlation analysis, $r_p$ lies always
within $12\%$ of that of model 1 for Fe/H and within $14\%$ of that
of model 1 for $\Gamma_{\rm norm}$. This also explains why our result contradicts
with that of HCT10 inspite of overall good agreement between predicted
number -- according to HCT10, the GCs with descending order of ranks are as follows Terzan~5, 47~Tuc, M\,28, NGC~6440, NGC~6441,
NGC~6752, M\,13, M\,5, M\,3, which is different from what we obtain.

\begin{table}\footnotesize
\caption{Results of different statistical correlation tests between the predicted number of pulsars with various GC parameters.}
\begin{tabular}{|l|r|r|r|r|r|r|} \hline \hline 
GC & \multicolumn{2}{|c|}{Pearson} & \multicolumn{2}{|c|}{Spearman} & \multicolumn{2}{|c|}{Kendall} \\
property & $rp$ & $P_{rp}$ & $rs$ & $P_{rs}$ & $\tau$ & $P_{\tau}$ \\ \hline
$\Gamma_{\rm norm}$ & 0.60 & 0.07 & 0.80 & 0.02 & 0.67  & 0.01  \\  
Fe/H & 0.51 & 0.13 & 0.44 & 0.18 & 0.31 & 0.21 \\
$d_{\rm gcen}$ & --0.64 & 0.04 & --0.69 & 0.04 & --0.54  & 0.03 \\
$\log(\rho_c)$  & 0.64 & 0.04 & 0.62 & 0.06 & 0.45 & 0.07 \\
 $c$ & --0.22 & 0.55 & --0.24 & 0.48 & --0.16 & 0.52 \\
$\log(t(r_c))$ & --0.29 & 0.41 & --0.51 & 0.13 & --0.34 & 0.17 \\
$M_{\rm GC}$ & --0.25 & 0.48 & --0.23 & 0.49 & --0.22 & 0.37 \\
$v_{c}$ & 0.31 & 0.38 & 0.58 & 0.08 & 0.31 & 0.21 \\
$r_c$ & --0.60 & 0.07 & --0.52 & 0.12 & --0.40 & 0.10 \\
\hline \hline
\end{tabular}
\label{tb:stat}
\end{table}

\section{Conclusions} 
\label{sec:conclu} 

We have modeled the observed luminosity distribution of millisecond
pulsars in globular clusters as the brighter tail of a parent
distribution. We have found that either a log-normal or a power-law
distribution can be used as the parent distribution. We have demonstrated
that a wide range of possible luminosity functions are compatible with
the data, and that log-normal distribution functions provide a better
match to the data than the traditionally favored power-law
distributions. In the light of these results, we conclude that
there is currently no need to assume that the
luminosity function for cluster pulsars is any different than that of
pulsars in the Galactic disk found by FK06. Based on this result, it
is quite possible that all pulsars follow similar luminosity
distribution irrespective of their positions or recycling history.

Contrary to earlier claims by HCT10, we find no evidence for a
significant correlation between the inferred numbers of radio pulsars
in GCs and either metallicity or stellar encounter rate. No significant
correlations were found against other cluster parameters either. 
Despite the lack of any obvious correlations found among this
sample of 10 GCs, it is of great interest to perform an analysis
using a much larger sample of clusters. Further
constraints may be possible by incorporating observations of the
diffuse gamma-ray flux, though this approach is complicated by model
dependencies in gamma-ray efficiency and the radio/gamma-ray beaming
fraction.

One key difference
between the two luminosity functions we have not commented on thus far
is shown in Tables \ref{tb:predlnorm} and \ref{tb:predplawLmax}
 by the tabulated parameter
$N_{10}$, the sum of the population estimates across all 10 GCs. As
can be seen, the power-law models predict a systematically larger
parent population than for the log-normal distribution (i.e. $N_{10}$ in
the range 1600--3800 compared to 350--700). This observation implies
that the power-law distributions require larger birth rates over the
log-normal models by a factor of 2--10. Although we shall defer
a detailed population size analysis to a future paper, containing
population estimates for more GCs, a simple scaling of these numbers
to all 150~GCs currently known implies a population range for potentially
observable recycled pulsars of  5000--11000
pulsars in the log-normal models versus 24000--57000 for the power law
models. Assuming a recycled pulsar lifetime of $\sim 10^{10}$~yr, and
a mean beaming fraction of 50\%, the
implied birth rate of this population is at least $10^{-6}$~yr$^{-1}$
over all Galactic GCs. 
Recent results concerning the low-mass X-ray binary (LMXB) population
in GCs \citep[see][for reviews]{hei11, pool10} suggest that there are of order
200 LMXBs in Galactic GCs. Assuming a typical LMXB lifetime of order
$10^8$~yr \citep{kn88}, the implied birthrate is comparable to our
rough estimates for the recycled pulsars, provided that the pulsar
population estimates are closer to the ranges suggested by the log-normal
models. 

We consider this study to be the first step towards a more comprehensive
analysis of the pulsar content of GCs. 
More detailed studies of the pulsar luminosity functions 
which better account for the selection effects and detection 
issues in the various radio surveys are still needed to further probe
all these issues. In particular, a search for correlations
between cluster parameters and the pulsar content beyond
the small sample of 10 GCs considered here is needed to better
understand this diverse population of neutron stars.

\section*{Acknowledgements}

We thank the anonymous referee and Craig Heinke for useful comments on
an earlier version of the manuscript. This work was supported by a
Research Challenge Grant to the WVU Center for Astrophysics by the
West Virginia EPSCoR foundation, and also from the Astronomy and
Astrophysics Division of the National Science Foundation via a grant
AST-0907967.

\end{document}